\documentclass[lettersize,journal]{IEEEtran}

\usepackage{amsmath}
\usepackage{graphicx}
\usepackage{fancyhdr}
\usepackage{enumitem}
\usepackage{url}
\usepackage{cite}
\usepackage[numbers,sort&compress]{natbib}
\usepackage{bbding}
\usepackage{booktabs}
\usepackage{amsfonts,amssymb}
\usepackage{bm}
\usepackage{multirow}
\usepackage{diagbox}
\usepackage{soul} 
\usepackage{makecell}
\usepackage{color}
\usepackage{pifont}
\usepackage{subfigure}
\usepackage[ruled, lined, linesnumbered, commentsnumbered, longend]{algorithm2e}
\usepackage[dvipsnames]{xcolor}
\usepackage{tikz,hyperref}
\definecolor{lime}{HTML}{A6CE39}
\DeclareRobustCommand{\orcidicon}{%
    \begin{tikzpicture}
    \draw[lime, fill=lime] (0,0) 
    circle [radius=0.16] 
    node[white] {
   {\fontfamily{qag}\selectfont \tiny ID}};    \draw[white, fill=white] (-0.0625,0.095) 
    circle [radius=0.007];    \end{tikzpicture}
    \hspace{-2mm}}
\foreach \x in {A, ..., Z}{%
    \expandafter\xdef\csname orcid\x\endcsname{\noexpand\href{https://orcid.org/\csname orcidauthor\x\endcsname}{\noexpand\orcidicon}}
    }

\usepackage{listings}
\definecolor{codegreen}{rgb}{0,0.6,0}
\definecolor{codegray}{rgb}{0.5,0.5,0.5}
\definecolor{codepurple}{rgb}{0.58,0,0.82}
\definecolor{backcolour}{rgb}{0.98,0.98,0.98}

\lstdefinestyle{mystyle}{
    backgroundcolor=\color{backcolour},   
    commentstyle=\color{codegreen},
    keywordstyle=\color{magenta},
    numberstyle=\tiny\color{codegray},
    stringstyle=\color{codepurple},
    basicstyle=\ttfamily\footnotesize,
    breakatwhitespace=false,         
    breaklines=true,                 
    captionpos=b,                    
    keepspaces=true,                 
    showspaces=false,                
    showstringspaces=false,
    showtabs=false,                  
    tabsize=2
}
\lstdefinelanguage{p4}
{morekeywords={table, action, Register, RegisterAction}, 
sensitive=false, 
}

\def\semichecked{\checkmark\!\!\!\raisebox{0.4 em}{\tiny$\smallsetminus$}}

\begin{document}
\bstctlcite{IEEEexample:BSTcontrol} 

\title{DeviceRadar: Online IoT Device Fingerprinting \\ in ISPs using Programmable Switches}

\author{
Ruoyu~Li\orcidA{},~\IEEEmembership{Graduate Student Member,~IEEE,}
Qing~Li\orcidB{},~\IEEEmembership{Senior Member,~IEEE,}
Tao~Lin,\\
Qingsong~Zou,
Dan~Zhao,
Yucheng~Huang,
Gareth~Tyson,
Guorui~Xie,
Yong~Jiang,~\IEEEmembership{Member,~IEEE}

\thanks{Ruoyu Li, Tao Lin, Yucheng Huang, Qingsong Zou, Guorui Xie and Yong Jiang are with Shenzhen International Graduate School, Tsinghua University, Shenzhen, China (e-mail: \url{liry19@mails.tsinghua.edu.cn}; \url{lint22@mails.tsinghua.edu.cn}; \url{huangyc20@mails.tsinghua.edu.cn}; \url{zouqs21@mails.tsinghua.edu.cn}; \url{xgr19@mails.tsinghua.edu.cn}; \url{jiangy@sz.tsinghua.edu.cn}).}

\thanks{Qing Li and Dan Zhao are with the Department of Mathematics and Theories, Peng Cheng Laboratory, Shenzhen, China (e-mail: andyliqing@gmail.com; zhaod01@pcl.ac.cn).}

\thanks{Gareth Tyson is with the Department of Electronic \& Computer Engineering, Hong Kong University of Science and Technology, Guangzhou, China (e-mail: gtyson@ust.hk).}

\thanks{\textit{Corresponding author: Qing Li}.}

}

\markboth{Journal of \LaTeX\ Class Files,~Vol.~14, No.~8, August~2021}%
{Shell \MakeLowercase{\textit{et al.}}: A Sample Article Using IEEEtran.cls for IEEE Journals}


\maketitle

\begin{abstract}
Device fingerprinting can be used by Internet Service Providers (ISPs) to identify vulnerable IoT devices for early prevention of threats.
However, due to the wide deployment of middleboxes in ISP networks, some important data, e.g., 5-tuples and flow statistics, are often obscured, rendering many existing approaches invalid. It is further challenged by the high-speed traffic of hundreds of terabytes per day in ISP networks.
This paper proposes DeviceRadar, an online IoT device fingerprinting framework that achieves accurate, real-time processing in ISPs using programmable switches.
We innovatively exploit ``key packets'' as a basis of fingerprints only using packet sizes and directions, which appear periodically while exhibiting differences across different IoT devices.
To utilize them, we propose a packet size embedding model to discover the spatial relationships between packets. Meanwhile, we design an algorithm to extract the ``key packets'' of each device, and propose an approach that jointly considers the spatial relationships and the key packets to produce a neighboring key packet distribution, which can serve as a feature vector for machine learning models for inference.
Last, we design a model transformation method and a feature extraction process to deploy the model on a programmable data plane within its constrained arithmetic operations and memory to achieve line-speed processing. 
Our experiments show that DeviceRadar can achieve state-of-the-art accuracy across 77 IoT devices with 40 Gbps throughput, and requires only 1.3\% of the processing time compared to GPU-accelerated approaches.
\end{abstract}


\setlength{\textfloatsep}{.8\baselineskip}
\setlength{\intextsep}{.8\baselineskip}

\section{Introduction}
Recent years have witnessed the rapid deployment of the Internet of Things (IoT). 
Meanwhile, insecure IoT devices are considered to remain one of the major concerns in networks over the foreseeable future~\cite{Bitag}. 
As IoT devices typically lack sufficient security protection, they have become a key target of botnet malware (e.g., Bashlite~\cite{bashlite}, Mirai~\cite{mirai}).
This situation makes Internet Service Providers (ISPs) increasingly concerned with vulnerable IoT devices connected to their networks. Take Mirai as an example: the compromised IoT devices were once used to launch large-scale DDoS attacks over 600 Gbps, wasting massive resources and sabotaging core ISP services such as DNS~\cite{mirai}.

To avoid being penetrated by \textcolor{black}{malicious} IoT devices, \textit{device fingerprinting} can be utilized by network administrators as a defense, which can identify the types of devices within its domain. 
\textcolor{black}{
This knowledge can then facilitate flexible precautions against high-risk devices (e.g., with explicit vulnerabilities on CVE~\cite{cve}) \textit{before} they get compromised or involved in malicious activities, e.g., by throttling, quarantining, limiting access to core infrastructures, or informing the users of the high-risk devices existing in their residence~\cite{pinpointing,broadcasting}.
Further, as the amount of IoT devices is surging (about 127 new devices per second in 2017~\cite{iot_increase}), it becomes necessary to do this with faster processing speed. For example, in 2020, researchers observed over 300 million login attempts by Mirai on 7500 IoT honeypots in 6 weeks~\cite{mirai_2020} (on average, each host receives 40 attacks per hour), suggesting that vulnerable IoT devices are in danger of being compromised at any time. 
Online device fingerprinting can mitigate this situation by timely identifying vulnerable devices and then preventing potential malicious activities in advance. 
For example, if an LG SuperSign TV known to be vulnerable to CVE-2018-17173 is detected, it can be protected by a rule of checking HTTP requests to port 9080 that exploit remote code execution.
Such a method can also provide prior knowledge to other security systems like IDS/IPS to increase their efficacy.
}




However, achieving effective and efficient online device fingerprinting is challenging in ISP networks for two reasons. 
\emph{First}, most existing works only investigate ideal local networks (LANs), where traffic can be easily separated by each device~\cite{sentinel,spy,traffic_shaping,peek-a-boo}.
In reality, traffic in ISP networks could originate from diverse gateways and middleboxes that hamper traffic analysis, e.g., Network Address Translation (NAT) gateways, Virtual Private Network (VPN) gateways, and The Onion Routers (Tor) nodes.
Due to traffic fusion and possible encryption and encapsulation, popular features used by existing approaches, including 5-tuples and various traffic statistics (e.g., packet counts, flow duration), could become unavailable or unreliable. \emph{Second}, today's ISP networks need to handle hundreds of terabytes of traffic per day.
Even if ISPs have sufficient server resources to accelerate the identification process, the overhead of data exchange is severe, such as data from network devices (i.e., data plane) to servers (i.e., control plane) and flow rules in the opposite direction.
The delay that such communications introduce could be over tens of seconds using conventional Software Defined Networking (SDN), which is non-trivial considering how much traffic must be forwarded.
Achieving real-time device fingerprinting is necessary yet challenging in the face of such high throughput.

In this paper, we propose DeviceRadar, a novel device fingerprinting framework that achieves high-speed processing in ISP networks with middleboxes. 
Through analyzing the traffic of multiple IoT devices, we observe that they periodically generate some bursts of traffic with cloud servers or IoT hubs, e.g., for data synchronization. These bursts typically contain a set of ``key packets'', which have stable sizes and are likely to appear in neighboring locations. Since these key packets exhibit differences among devices and can be characterized only by packet sizes and directions, which are reliable in middlebox scenarios, in DeviceRadar, we utilize them as a basis of fingerprints for accurate device identification. However, due to inevitable packet loss, disorder and retransmission in ISP networks, simply matching the sequences of these packets to the online traffic might fail. 
For example, retransmission, duplicate ACK and out-of-order packets account for 5.4\% in one-day trace of WIDE backbone~\cite{mawi}.

To utilize key packets for fingerprinting, we first propose a \textit{packet embedding} model that brings packet sizes to a high-dimensional space where correlated packets are in closer positions. With this model, we can predict the probability of key packets appearing in the neighboring position of given packets using the spatial distances between the embeddings. 
The result of predicted probabilities does not require precise matching of packet sequences like other signature-based approaches (e.g.,~\cite{pingpong}); 
instead, it forms a feature vector as fingerprints for simple ML classifiers to identify the target devices. 
As it only uses packet sizes and directions, DeviceRadar can be applied to handling complex middlebox scenarios. 

To address the challenge of runtime overhead, we exploit \textit{P4 programmable switches}~\cite{p4}, which open up the possibility of in-network computing. As P4 switches suffer from arithmetic operation limitations (e.g., not supporting loops, division or float-point operations), we design a method of transformation from our models (i.e., packet embedding, ML classifiers) to P4 match-action tables to realize line-rate ML inference within the pipeline of packet processing. Moreover, to mitigate the memory constraint of P4 switches (e.g., TCAM, SRAM), we develop an incremental feature construction process using P4 stateful registers for online traffic. With these designs, we manage to bypass the restrictions of programmable switches and deploy DeviceRadar fully on the data plane, which can achieve the line-rate processing speed for online use.

We prototype DeviceRadar on a physical P4 switch.
For evaluation, we construct a real-world IoT testbed and collect a three-month traffic dataset, and use three public IoT datasets as benchmarks and a backbone trace as background traffic. 
We demonstrate two common but challenging middlebox scenarios: NATs and VPNs.
The experiments show that DeviceRadar can achieve high identification accuracy across 77 IoT devices with 40 Gbps throughput and only 1.3\% of the processing time compared to the GPU-accelerated methods. 

The contributions of this paper are summarized as follows.
We present 
1) A novel in-network device fingerprinting framework for ISP networks, which achieves high accuracy, high throughput and low processing time; 
2) A packet embedding model that predicts the packet sizes in the neighboring position, which promotes the efficacy of traffic analysis; 
and
3) A prototype of DeviceRadar on physical hardware, and a real-world IoT testbed for realistic evaluation.

\section{Background and Related Work}

\subsection{IoT Device Fingerprinting by ISPs}

IoT device fingerprinting identifies a specific set of device types by passively sniffing network traffic.
\textcolor{black}{
Many IoT device fingerprinting approaches have been proposed for various scales of networks, such as public Wi-Fi networks~\cite{broadcasting}, wireless sensor networks~\cite{passive,GTID}, and home networks~\cite{sentinel,homesnitch,hanzo}. Some studies describe their works from the view of an adversary for privacy sniffing~\cite{pingpong,spy,no_castle}. 
Different layers of information have been used for device identification. For example, Radhakrishnan et al. introduce a technique that can fingerprint types of wireless devices by utilizing physical-layer information~\cite{GTID}. 
Franklin et al. develop a wireless device driver fingerprinting method based on the data link layer~\cite{passive}. 
In contrast, this paper focuses on IoT device fingerprinting by ISPs who can only monitor network traffic on the links and network devices with the ISP domain. Note, sniffing wireless IoT packets over the air is impractical for ISPs, as this can only be done close to the signal emitters (e.g., at the WiFi access point). As such, prior techniques based on lower-layer information are unavailable for our scenario.
}

\textcolor{black}{
Though there have been works on IoT device identification for large-scale networks by \textit{offline} analysis~\cite{haystack,iotfinder}, this paper mainly explores its benefit to \textit{online} network management -- given the insecurity of IoT, high-risk devices in the network can be timely pinpointed so that their communications can be constrained in advance according to their known vulnerabilities. The identification result can also facilitate downstream systems (e.g., anomaly detection/prevention) by providing prior knowledge of device labels, simplifying their working logic.
}
We argue that IoT device fingerprinting by ISPs in an online manner should satisfy the following requirements:

\noindent \textbf{R1) High accuracy.} The target IoT devices should be precisely identified even if the non-IoT traffic (i.e., background traffic) might dominate the volume of data.

\noindent \textbf{R2) Real-time processing.} The identification result should be timely so that prompt actions could be taken before substantial malicious traffic has flowed into the network.


\noindent \textbf{R3) High throughput.} Given the high rate of ISP networks (e.g., tens of Gbps), an online system should quickly process and forward a huge amount of traffic data.

However, achieving the above requirements faces two major challenges:

\subsubsection{Middleboxes}
\label{sec:middlebox}
Middleboxes are devices widely deployed in autonomous systems and across various networks including ISPs~\cite{middlebox}, such as firewalls, load balancers, DPI boxes, NATs, VPNs, onion routers.
As they can not only inspect but manipulate traffic, the difficulty of accurate identification (i.e., R1) has markedly increased as many useful traffic features can be obscured. For IoT connections, we mainly consider two types of middleboxes: NATs and VPNs. Table~\ref{tab:nat_vpn} summarizes the traffic features modified by NATs and VPNs.

\textbf{NATs} are commonly used to relieve the exhaustion of public IP addresses. As the source address is multiplexed, host-level traffic statistics (e.g., statistics of packet sizes and inter-arrival time per host) can be fused by the traffic of multiple devices. In addition, the typical Network Address Port Translation (NAPT) maps the source port of devices to a new port for address translation. These characteristics aggravate the difficulty of device fingerprinting for ISPs.

\textbf{VPNs} have been commercialized for decades both as standalone products and as integrated parts of firewall products (e.g.,~\cite{paloalto}). 
They are usually located on the boundaries of networks to provide secure data transmission and remote access. When traffic passes through VPNs, the source and the destination are replaced by a uniform tunnel between two VPN endpoints. 
Furthermore, most mainstream VPN protocols can encrypt and encapsulate the original packets, rendering most of the header fields unavailable. For example, the entire L3 packet is encrypted and encapsulated in the payload of an IPSec ESP packet; for SSL/TLS VPNs (e.g., OpenVPN), the destination port is set to 1194 and the layer 4 is encapsulated with a new TCP/UDP header.

In summary, we argue that only packet sizes and packet directions are reliable features that can be generalized across different middlebox scenarios.

\begin{table}[htbp]
    \centering
    \footnotesize
    \caption{Traffic features modified by NATs and VPNs.}
    \begin{tabular}{c|c|cc}
    \toprule
    \multirow{2}{*}{Feature} & \multirow{2}{*}{NAT} & \multicolumn{2}{c}{VPN} \\
    ~ & ~ & IPSec & SSL/TLS \\
    \midrule
    src-IP & NAT's external IP & \multicolumn{2}{c}{VPN's external IP} \\
    dst-IP/domain & - & \multicolumn{2}{c}{VPN remote endpoint} \\
    src-port & translated & encrypted & VPN's src-port \\
    dst-port & - & encrypted & service port \\
    L4 protocol & - & encrypted & encapsulated \\
    host-level stats & mixed & mixed & mixed \\
    flow-level stats & - & mixed & mixed \\
    \bottomrule
    \end{tabular}
    \label{tab:nat_vpn}
\end{table}

\subsubsection{Runtime Overhead}
\begin{figure}[htbp]
    \centering
    \subfigure[Raw traffic by port mirroring]{
    \begin{minipage}{0.54\linewidth}
    \centering
    \includegraphics[width=\textwidth]{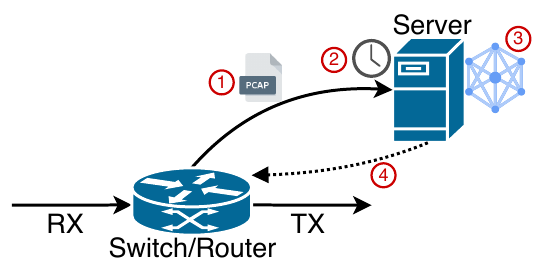}
    \end{minipage}
    }
    \subfigure[SDN with OpenFlow]{
    \begin{minipage}{0.38\linewidth}
    \centering
    \includegraphics[width=\textwidth]{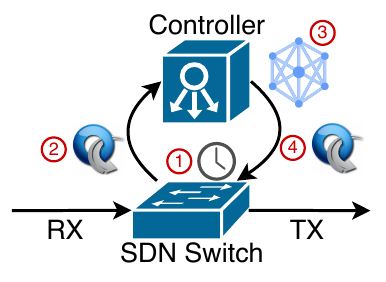}
    \end{minipage}
    }
    \caption{Runtime overhead of two off-path deployments.}
    \label{fig:deploy}
\end{figure}

Achieving device fingerprinting as an online system is challenging in an ISP network in view of its high-speed traffic. 
We illustrate two common deployments in Fig. \ref{fig:deploy}. One approach is to copy the raw traffic by port mirroring to a server that runs the device identification algorithm (e.g.,~\cite{secret,denat,byteiot,pinpointing}). Though this approach is practical for small networks like home networks, the generic server is difficult to process the high-throughput online traffic of ISP networks that can reach tens of gigabytes per second (i.e., R3). 

Another approach is based on the SDN paradigm, where an SDN switch processes the traffic and uses OpenFlow to send traffic statistics to and accept flow rules from the controller (e.g.,~\cite{sentinel,deft}). This approach can obviously reduce the load of the controller server. However, it still suffers from the inherent drawback of the long control loop in off-path deployments, making the real-time processing unrealistic (i.e., R2). Specifically, a round of processing consists of:

\noindent 1) Time window to collect traffic features and calculate statistics for one inference (\ding{172}), typically seconds to minutes; 

\noindent 2) Communication latency, including statistics uploading (\ding{173}) and flow rule issuing (\ding{175}), typically tens of milliseconds and even more under high load; 

\noindent 3) Inference time that the identification algorithm consumes, typically tens of milliseconds (\ding{174}).

In summary, these deployments introduce much runtime overhead that may increase as networks grow in capacity. For security in ISP networks, the latency of seconds to minutes before the installation of the defense rules can drastically devalue the identification of vulnerable devices, since a huge amount of potentially malicious traffic may have been forwarded.

\subsection{Limitations of Existing Work}
\label{sec:limits}

According to the key features and techniques, we categorize existing methods into three types as explained in Table~\ref{tab:related_work}.

\textbf{Signature-based methods.} 
Traditional signatures, including MAC addresses/OUI~\cite{oui} and DHCP messages~\cite{fingerbank}, achieve real-time device identification in a LAN but are mostly infeasible for ISPs. Many studies use the set of destination IP addresses and domains via DNS queries as distinguishable IoT signatures~\cite{spy,no_castle,traffic_shaping,haystack,iotfinder,haystack_workshop}. Though these approaches are effective, it has been pointed out that they can be invalidated if the DNS is encrypted or different vendors use the same public cloud services~\cite{pinpointing,byteiot}. Besides, collecting the full set of DNS queries needs a long time window and is often hit-and-miss.
For instance, we observe that a TP-Link camera in our testbed only has one DNS query per hour on average; among its 11 domains, 5 are only queried immediately after network connection and never queried again. 
It naturally hinders timely and reliable device identification (i.e., R1 \& R2). 
Trimananda et al.~\cite{pingpong} discover a packet-level signature that contains pairs of packets with predictable lengths.
However, it only supports device fingerprinting over TCP connections and cannot handle UDP-based devices, which limits its practical usage.

\textbf{ML-based methods.} 
Machine learning (ML) algorithms enable device fingerprinting to utilize multi-dimensional feature vectors for inference, including packet header fields~\cite{sentinel,lumos,denat} (e.g., protocol, port, timestamp, TCP flag, option) and traffic statistics~\cite{passive,GTID,darkside,deft} (e.g., mean/maximum/minimum/variance of count/size/inter-arrival time/duration). 
For example, Meidan et al. propose an approach using header features and the LGBM algorithm~\cite{denat}. DarkSide~\cite{darkside} adopts 16 traffic statistics to infer IoT devices. Compared to signature-based methods, these approaches are more flexible and usually do not rely on specific protocols. 
However, due to the middleboxes that obscure header fields and fuse the IoT traffic with other non-IoT traffic, many of these approaches can be inaccurate in ISP networks (i.e., R1).
Besides, obtaining traffic statistics may require a long time window, such as the default 30 minutes for active flows in NetFlow, which greatly reduces the real-time nature of the detection result (i.e., R2).

\textbf{DL-based methods.} More recent approaches adopt state-of-the-art deep learning (DL) algorithms that can better discern the sequential patterns among packets. For example, Dong et al. propose HomeMole based on bidirectional LSTM to leverage the temporal relationship between packets, which shows to be effective in NAT scenarios but relatively weak in VPN scenarios~\cite{secret}. 
Ma et al. design a system via spatial-temporal traffic fingerprinting and a CNN to identify IoT devices hidden behind NATs~\cite{pinpointing}. 
Though achieving better accuracy, this type of method may not be suitable for online use in high-speed ISP networks, given that their inference time (e.g., milliseconds on GPUs) is much greater than the average packet inter-arrival rate (e.g., microseconds).
In addition, they need a long window to better capture the spatial/temporal relationships, such as several minutes in~\cite{pinpointing}, which further degrades the timeliness of the results (i.e., R2).

Besides, none of the previous works consider the requirement of throughput in realistic ISP networks (i.e., R3). In summary, 
\textcolor{black}{
there is no prior work that addresses the need for online device fingerprinting in ISP networks which meets the aforementioned requirements.
}

\begin{table}[h]
    \caption{Existing methods versus ours in ISP networks.}
    \label{tab:related_work}
    \footnotesize
    \centering
    \begin{tabular}{c|c|c|c|c}
    \toprule
    Method & \makecell[c]{Key feature \\ \& technique} & \makecell[c]{Accurate w/ \\ middlebox} & Real-time & \makecell[c]{High- \\ throughput} \\
    \hline
    \multirow{3}{*}[-10pt]{Signature} & \makecell[c]{MAC/DHCP \\ \cite{oui,fingerbank}} & \XSolidBrush & \Checkmark & \XSolidBrush \\
    \cline{2-5}
    ~ & \makecell[c]{IP/domain \\ \cite{spy,no_castle,haystack,iotfinder}} & \scalebox{1.4}{\semichecked} & \XSolidBrush & \XSolidBrush \\
    \cline{2-5}
    ~ & \makecell[c]{packet pair \\ \cite{pingpong}} & \scalebox{1.4}{\semichecked} & \XSolidBrush & \XSolidBrush \\
    \hline
    \multirow{2}{*}[-5pt]{ML} & \makecell[c]{header field \\ \cite{sentinel,lumos}} & \XSolidBrush & \XSolidBrush & \XSolidBrush \\
    \cline{2-5}  
    ~ & \makecell[c]{traffic statistic \\ \cite{passive,GTID,darkside,deft}} & \XSolidBrush & \XSolidBrush & \XSolidBrush \\
    \hline
    DL & \makecell[c]{spatial/temporal \\ relationship \\ \cite{secret,pinpointing}} & \scalebox{1.4}{\semichecked} & \XSolidBrush & \XSolidBrush \\   
    \hline
    \makecell[c]{Ours} & \makecell[c]{embedding + \\ P4 switch} & \Checkmark & \Checkmark & \Checkmark \\
    \bottomrule
    \end{tabular}
\end{table}

\subsection{Programmable Data Plane and In-network Intelligence}
\label{sec:dataplane}

Recent research has extended the conventional SDN architecture from the programmable control plane to the programmable data plane. A programmable switch enables packet processing in arbitrary formats and protocols defined by users.
As such, many tasks like load balancing~\cite{silkroad}, RTT measurements~\cite{rtt} and firewalling~\cite{p4guard} can be offloaded directly to the data plane. Exploiting the on-path deployment with microsecond-level processing latency (i.e., R2) and Tbps-level throughput (i.e., R3) on programmable switches sheds new light on the implementation of online network tasks.

\begin{figure}[tp]
    \centering
    \includegraphics[width=.8\linewidth]{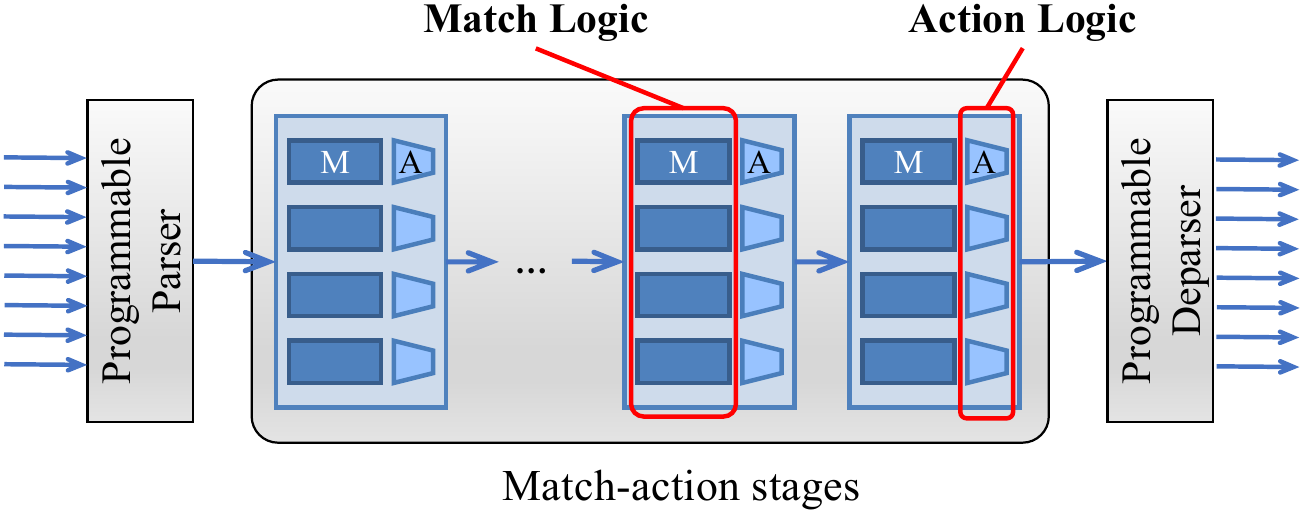}
    \caption{Protocol Independent Switch Architecture (PISA).}
    \label{fig:pisa}
\end{figure}

This versatility and efficiency are realized by a programmable Application-Specific Integrated Circuit (ASIC) for networking (e.g., Intel Tofino \cite{tofino}).
It follows the Protocol Independent Switch Architecture (PISA), as illustrated in Fig.~\ref{fig:pisa}. PISA consists of a programmable parser, a series of match-action stages and a programmable deparser. The match logic uses a mix of SRAM and TCAM for lookup tables, registers, hash tables and other data structures.
The action logic uses ALUs for standard boolean and arithmetic operations, header modifications, hashing operations, etc. A network-specific programming language, P4~\cite{p4}, is available to write and load custom programs for processors of PISA.

However, to guarantee high-speed processing, most P4 switches are designed with the constraints of 
1) \textit{operations} that only support simple instructions like integer additions and bit shifts, but do not support loop, division or floating-point operations; 
and 
2) \textit{resources} that have limited match-action stages (e.g., 4 pipelines of 12 stages in Tofino) and memory. 
It means that most ML/DL models are difficult to be executed on P4 switches, implying the dilemma between high accuracy (i.e., R1) and low runtime overhead (i.e., R2 \& R3).
Recently, Xiong et al. first explore the potential of mapping specific ML models, including decision tree, Naïve Bayes, K-Means and SVM, to match-action pipelines in a P4 switch for deployment~\cite{iisy}. Among these models, tree-based models (e.g., decision trees) are more suitable as their rule-based decision process naturally aligns with the match-action pipelines. Several studies have proposed in-network intelligence solutions using tree-based models~\cite{in-network_classification,mousika,pforest}.

Nevertheless, prior studies~\cite{tabular} have shown that tree-based models may suffer from accuracy problems, as they are not good at learning spatial/temporal relationships within sequential data, which is the key information used by some approaches~\cite{secret,pinpointing} to resolve the traffic mixing issue of middleboxes. Thus, a model that uses reliable traffic features to unearth the relationship among packets and can fit the obtained relationship to tree-based models for deployment is needed.

\section{Threat Model}
This paper focuses on IoT device fingerprinting from the view of an ISP.
We propose a security and management tool for identifying the types of IoT devices within its administration domain. 
\textcolor{black}{
We use the term ``device type'' to indicate devices with the same manufacturer and functions (e.g., ``Xiaomi-plug''). 
From the perspective of security, this granularity is sufficient to pinpoint the behavior profile and vulnerability of these devices (since they typically share a similar set of firmware).
Note that there is a difference between our objective and individual device identification which is out of the scope of this paper.
}

The adversary considered in this paper is the owner of IoT botnet who seeks to attack, compromise and control large numbers of IP-enabled IoT devices. The existence of compromised IoT devices poses a threat to an ISP's infrastructure, services and clients, e.g., DDoS attacks on DNS services, high-profile websites and even national Internet infrastructure in the incident of Mirai in 2016~\cite{mirai}.
We assume that the adversary targets a limited number of device types. For example, Mirai typically only targets IP cameras, DVRs and routers~\cite{mirai}.

As most consumers do not have expert knowledge to prevent such attacks, an ISP must take the responsibility to identify those vulnerable IoT devices before they are exploited by attackers. Unlike the device fingerprinting in a LAN~\cite{sentinel,spy,traffic_shaping,peek-a-boo}, in ISP networks (where traffic through middleboxes is prevalent) we assume that traffic from different devices cannot be reasonably separated by IP addresses (e.g., the upstream comes from a NAT gateway) nor by 5-tuples (e.g., the upstream comes from a VPN gateway). 
For the sake of generality, this paper assumes that an IP address could carry the traffic of multiple devices, including both IoT and non-IoT traffic.
Further, in most cases, we assume IoT traffic is sparse compared to background traffic. Except inside the LANs (e.g., homes, enterprises), the links that ISPs can monitor are assumed to be arbitrary, such as access/aggregation switches and core switches. Besides, ISPs can read the header fields of packets but do not tend to use DPI on payloads due to encryption or privacy policies. Lastly, we assume ISPs can arbitrarily obtain traffic samples of the target IoT devices by purchasing these devices and setting up a private testbed.

\begin{table}[b]
    \caption{Case study: two distinct devices demonstrate periodic and differential bursts of packet sizes and directions (minus sign indicates the direction from WAN to LAN).}
    \label{tab:xiaomi_plug}
    \small
    \centering
    \begin{tabular}{cc|cc}
    \toprule
    Xiaomi plug & Period & TP-Link camera & Period \\
    \midrule
    74, -74 & 30s & 167, -151, 66 & 30s \\
    111, -111, 60 & 40s & 321, 145, -145 & 5min \\
    175, -447, 191 & 5min & 251, -393, -123 & 15min \\
    543, -143, 431, -399 & 30min & 136, -871, -1486 & 15min \\
    \bottomrule
    \end{tabular}
\end{table}

\section{Overview}

\subsection{Observation of IoT Key Packets as Fingerprints}
\label{sec:key_pkt}

\textcolor{black}{
As discussed in Section~\ref{sec:limits}, accurate device fingerprinting in ISP networks is challenging due to information loss by middleboxes. 
In such circumstances, packet sizes and directions are the only reliable packet features available: packet sizes suggest data transmission load, and directionality differentiates packets sent from either IoT devices or IoT clouds, indicating communication patterns between clients and servers.
}

To further exploit the usefulness of the limited available features, we analyze the traffic of 14 IoT devices of different types and brands in our testbed, and observe some common characteristics among them. 
We present the results of a Xiaomi plug and a TP-Link camera collected over 10 days in Table \ref{tab:xiaomi_plug} as an example. 
We observe that bursts of packets with specific sizes appear periodically. For example, every 30 minutes four packets with the size and direction of 543, -143, 431 and -399 appear in the traffic of the Xiaomi plug, even though they may not follow the exact order due to packet reordering and retransmission. We refer to these packets the \emph{key packets} with respect to sizes and directions. 

We also observe that the key packets of different device types are different. It is in line with our further investigation on the IoT firmware development platforms (e.g., Xiaomi~\cite{xiaomi}, SmartThings~\cite{smartthings}), which reveals that these packets are typically for periodic \textit{property} synchronization, such as status, power and sensor data.
They vary among device types due to different properties (e.g., electricity usage for plugs, temperature for thermostats), formats (e.g., bool, int, float, string) and protocols (e.g., HTTP, MQTT). 
It means that, even if two devices may use some common public services, their key packets will not completely coincide as long as they are with either different manufacturers or different functions.
Besides, this type of periodic traffic always exists no matter if the device is in an idle or active state. This characteristic makes device identification possible at any given time. 
As such, we exploit the key packets as the basis of device fingerprints.

However, considering that packet disorder and retransmission are common in ISP networks, directly matching the arriving packets with the sequences of key packets like~\cite{pingpong} may often fail. Besides, these bursts of packets are easily hidden by high-speed background traffic, making them nearly impossible to be identified one by one.


\subsection{Overview of DeviceRadar}
\begin{figure}[t]
   \centering
   \includegraphics[width=\linewidth]{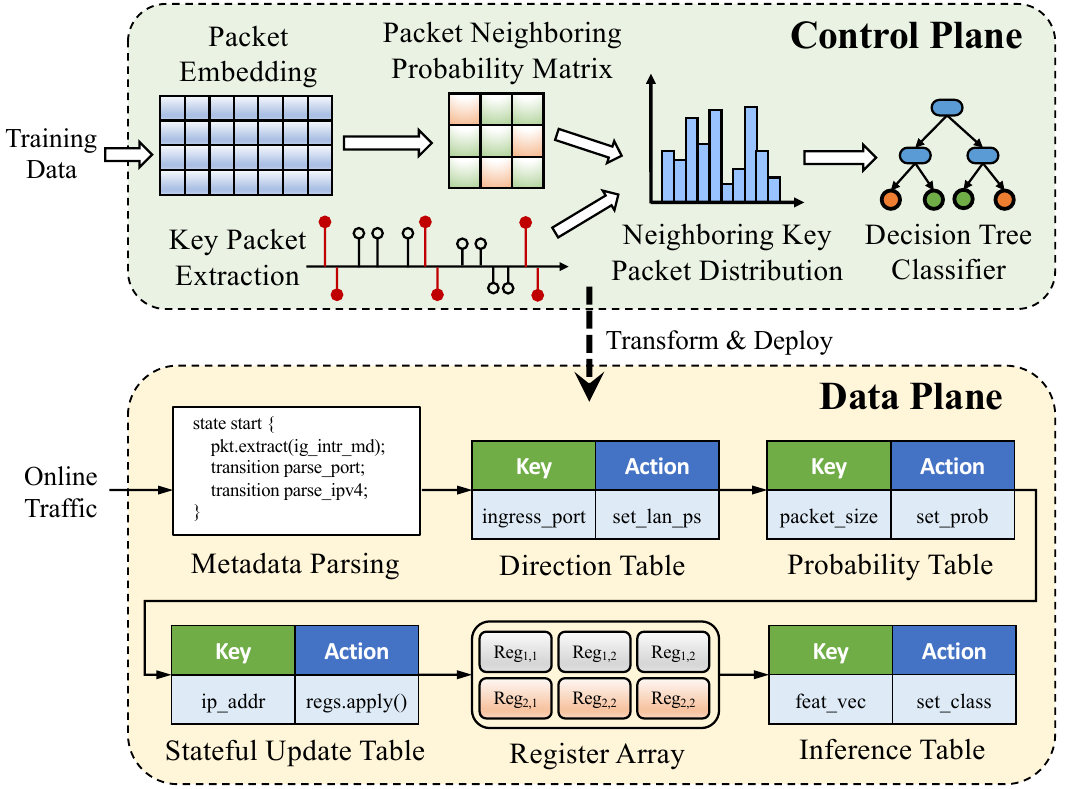}
   \caption{System architecture of DeviceRadar.}
   \label{fig:arch}
\end{figure}

We propose DeviceRadar, a novel framework for IoT device fingerprinting in ISP networks. 
\textcolor{black}{
To the best of our knowledge, it is the first implementation of online device identification that can handle complex traffic scenarios (like NATs and VPNs), achieve real-time processing, and support high throughput in one system.
}
Fig.~\ref{fig:arch} shows its architecture, which is built on an SDN paradigm divided into control plane and data plane.


\textbf{Control Plane.} The control plane is responsible for generating the fingerprinting models and issuing policies to the data plane.  
Inspired by our observation in Section~\ref{sec:key_pkt}, we exploit key packets as the base of IoT fingerprints, which just require packet sizes and directions, i.e., the limited reliable features in middlebox scenarios.
Specifically, we propose a packet size embedding model to discover the spatial relationships between packets, which can be further transformed into a packet neighboring probability matrix for each target device. Meanwhile, we design an algorithm to extract the ``key packets'' of each device that exhibit sufficient frequencies and periodic patterns. By combining the packet probability matrix and the list of key packets, DeviceRadar can obtain a neighboring key packet distribution for a certain time window, which can then be used as a feature vector for a decision tree classifier to accurately tell the existence of target devices.

\textbf{Data Plane.} 
At runtime, DeviceRadar eliminates the long communication loop with the control plane and can be completely offloaded to the data plane, which is the programmable switch ASIC with microsecond-level processing latency and Tbps-level throughput. It is because we find a way to transform the entire working flow and the models generated on the control plane to switch-compliant match-action pipelines and tables, and compile them as a P4 program that can be deployed on the data plane. To achieve this, we manage to overcome the restraint of operations and resources of the programmable switches during the transformation and deployment. With these efforts, DeviceRadar can fully take advantage of the high performance of switch ASICs and achieve real-time identification and high throughput.


\section{DeviceRadar Control Plane}
\label{sec:control_plane}
The control plane is responsible for the offline training of the models, 
including the construction of a \textit{packet embedding} model to amplify the correlation among packet sizes and directions within target IoT traffic, the acquisition of a \textit{packet neighboring probability matrix} which depicts the probability of certain packets co-occurring in bursts of traffic, the extraction of \textit{key packets} of target IoT devices, and finally the formation of \textit{neighboring key packet distribution} as feature vectors to train a \textit{decision tree classifier} for accurate inference and further online deployment on the data plane.


\subsection{Packet Embedding} 
\begin{figure}[htbp]
    \centering
    \includegraphics[width=.85\linewidth]{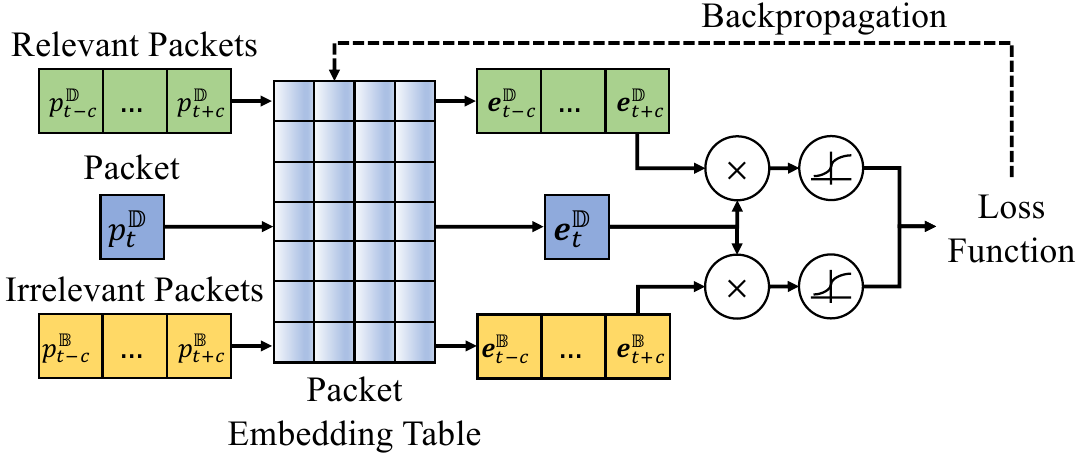}
    \caption{Training process of packet embedding model.}
    \label{fig:Pac2Vec}
\end{figure}

\textcolor{black}{
Our intuition is to automatically discover co-occurrence relationships between directional packet sizes in traffic. With this knowledge, we can predict surrounding packets being key packets cumulatively, assisting in detecting target IoT devices.
To fulfill this, we exploit \textit{word embedding} techniques from Natural Language Processing (NLP), and build a deep learning model for \textit{packet embedding}. 
The goal of word embedding is to transform a word into a high-dimensional encoding space where the context-dependent words have a higher similarity. Thus, the most likely neighboring words can be predicted by the spatial distance.
}
Formally, for a packet with size $s$ and direction $r$, we encode direction into packet size by:
\begin{equation}
    p = \begin{cases}
        s, & r=0 \\
        s + 1500, & r=1
    \end{cases},
\label{eq:dir_ps}
\end{equation}
where $r=0,1$ indicates the direction from LAN to WAN and from WAN to LAN, respectively. The embedding is essentially represented by a lookup table $A \in \mathbb{R}^{K\times d}$, where $K$ is the number of all possible directional packet sizes and $d$ is the dimension of the embedding. For IP packets, $K:=1500 + 1500$, where 1500 is the maximum transmission unit (MTU) of Ethernet. 
A packet\footnote{During the introduction to DeviceRadar, we use ``packets'' to refer to packet sizes encoded with directions for short.} $p$ can be converted into a $d$-dimensional embedding vector $\boldsymbol{e}$ by simply retrieving the $p$-th row of the array, i.e., $\boldsymbol{e} = A[p, :]$. For a target device $\mathbb{D}$, we use the pure traffic of the device, denoted by $P_\mathbb{D}$, and the skip-gram model to train the embedding, as illustrated in Fig.~\ref{fig:Pac2Vec}. 

The training process is in an unsupervised manner. For each packet $p^\mathbb{D}_t \in P_\mathbb{D}$, we consider the $c$ adjacent packets that arrive before and arrive later in a burst the \textit{relevant packets}, meaning that they are more likely to co-occur with the packet $p$.
\textcolor{black}{
For each relevant packet, we employ the unigram distribution to sample $k$ packets from the background traffic. We denote these \textit{irrelevant packets} as $P_\mathbb{B}$. The empirical equation to calculate the probability of being sampled is:
\begin{equation}
\mathrm{P}_{sample}(p_i^\mathbb{B}) = \begin{cases}
    \frac{f(p_i^\mathbb{B})}{\sum_{j=1}^{3000}f(p_j^\mathbb{B})}, & p_i^\mathbb{B} \notin \{p_{t-c}^\mathbb{D}, ..., p_{t+c}^\mathbb{D}\} \\
    0, & p_i^\mathbb{B} \in \{p_{t-c}^\mathbb{D}, ..., p_{t+c}^\mathbb{D}\}
    \end{cases},
\label{eq:sample}
\end{equation}
where $f(\cdot)$ is the packet frequency in $P_\mathbb{B}$. It means that, for all possible values of packets from 1 to 3000, packets with higher frequency in the background traffic are more likely to be sampled (first case), but packets equal to any of the relevant packets will not be selected (second case).} 
The training goal is to minimize the loss function:
\begin{equation}
    \mathcal{L}(\boldsymbol{e}^{\mathbb{D}}_{t}) = - \sum_{i=-c \atop i\neq 0}^{c} \sum_{j=1}^{k} (\log \sigma(\boldsymbol{e}^{\mathbb{D}}_{t} \cdot \boldsymbol{e}^{\mathbb{D}}_{t+i}) + \log \sigma(- \boldsymbol{e}^{\mathbb{D}}_{t} \cdot \boldsymbol{e}^{\mathbb{B}}_{j})), 
\label{eq:Pac2Vec}
\end{equation}
\textcolor{black}{
where $\sigma(\cdot)$ is the sigmoid function, and $\boldsymbol{e}$ denotes a $d$-dimensional embedding corresponding to a packet $p$.
For the embedding $\boldsymbol{e}^{\mathbb{D}}_{t}$ of $p^\mathbb{D}_t$, it means to maximize its similarity with the embeddings of its $(2c-1)$ relevant packets (i.e., the former term from $\boldsymbol{e}^{\mathbb{D}}_{t-c}$ to $\boldsymbol{e}^{\mathbb{D}}_{t+c}$), and minimize its similarity with the embeddings of the $k\cdot(2c-1)$ irrelevant packets (i.e., the latter term).
Consequently, the trained embedding tends to place the packets more likely to appear together in a closer position. We provide theoretical proof of this claim in the Appendix.
}

\subsection{Packet Neighboring Probability Matrix}
The packet embedding model can be further transformed to a packet probability matrix, which contains the probability of the neighboring packets, i.e., \textit{neighboring probability}. Formally, given a packet $p$, we denote the neighboring packet by variable $\mathcal{X}$ and the probability of $\mathcal{X}$ that appears near $p$ in the traffic of device $\mathbb{D}$ by $\mathrm{P}(\mathcal{X}|p;\mathbb{D})$. 
Since packets that are more likely to occur together tend to have higher similarity between their embedding vectors, we use the similarity in the packet embedding table to estimate this probability. Suppose there are $n$ possible packet sizes with frequencies not less than $\epsilon$ in the traffic data $P_\mathbb{D}$. We calculate the cosine similarities between the embedding of each packet and those of the other packets to obtain similarity matrix $S_\mathbb{D}$, which is given as: 
\begin{equation}
    S_\mathbb{D} = \begin{bmatrix}
        \frac{p_1 \cdot p_1}{|p_1||p_1|} & \frac{p_1 \cdot p_2}{|p_1||p_2|} & ... & \frac{p_1 \cdot p_{n}}{|p_1||p_n|} \\
        ... & ... & ... & ... \\
        \frac{p_{n} \cdot p_1}{|p_{n}||p_1|} & \frac{p_{n} \cdot p_2}{|p_{n}||p_2|} & ... & \frac{p_{n} \cdot p_{n}}{|p_{n}||p_{n}|} \\
    \end{bmatrix}.
\label{eq:dist_matrix}
\end{equation}

For cosine similarities in the range of $[-1, 1]$, those values in $S_\mathbb{D}$ lower than a threshold $\lambda \ge 0$ are truncated to zeros to obtain the \textit{packet neighboring probability matrix} $M_\mathbb{D}$:
\begin{equation}
    M_\mathbb{D}[i, j] = \begin{cases}
        S_\mathbb{D}[i, j] & S_\mathbb{D}[i, j] \ge \lambda \\
        0 & S_\mathbb{D}[i, j] < \lambda 
    \end{cases},
\label{eq:prob_matrix}
\end{equation}
as they indicate that the packets have highly different directions in the embedding space and thus are considered irrelevant. We empirically set $\lambda = 0.4$ in our implementation. 
Given a packet $p_i$ in the traffic of device $\mathbb{D}$, we denote the probability of a different packet $p_j$ appearing near $p_i$ by:
\begin{equation}
    \mathrm{P}(\mathcal{X}=p_j|p_i;\mathbb{D}) = M_\mathbb{D}[i, j].
\label{eq:neighbor_prob}
\end{equation}

\subsection{Key Packet Extraction}
Formally, we define key packets as packets satisfying the following characteristics: 
1) they frequently appear in a burst of device traffic; 
and
2) they have a stable period. 
We design a \textit{key packet extraction} algorithm as summarized in Algorithm~\ref{algo:key_packet}. 
First, we split the traffic by the tuple of destination IP, destination port and L4 protocol, which is likely for the same purpose.
For each subset of the traffic, we extract the bursts by a threshold of burst intervals $T_b$, and record each burst with its start timestamp and list of packets (line 4$\sim$11). Then we calculate the intervals between every two adjacent bursts and find if they have a stable period by the Coefficient of Variation $c_v$ of the burst intervals, which is the mean divided by the standard deviation and represents the normalized divergence of data (line 12$\sim$16). If these bursts have a sufficient frequency and a low $c_v$ (i.e., stable period), the packets in these bursts are extracted as a batch of the result, i.e., the key packets.

\begin{algorithm}[htbp]
\SetAlgoLined
\KwIn{Device packets $P_{\mathbb{D}}$ and their timestamps $I_{\mathbb{D}}$}
\SetKw{KwForIn}{in}
\SetKw{KwAnd}{and}
\SetKw{KwOr}{or}
\SetKw{KwFor}{for}
Initialize an empty key packet set $S$\;
Split traffic by the tuples of destinations\;
\For{$P$, $I$ \KwForIn $P_{\mathbb{D}}$, $I_{\mathbb{D}}$}{
Initialize an empty burst list $L_B$ and $t_{prev} \gets 0$\;
\For{$p$, $t$ \KwForIn $P$, $I$}{
\uIf{$t - t_{prev} > T_b$ \KwOr $t_{prev} = 0$}{
    $B \gets \{ts: t, pkts: []\}$\;
    $L_B.\text{append}(B)$;}
$B.pkts.\text{append}(p)$\;
$t_{prev} \gets t$;
}
Initialize an empty list $H$ of burst intervals\;
\For{$B$, $B_{next}$ \KwForIn $L_B$}{
    $H.\text{append}(B_{next}.ts -  B.ts)$\;
}
$c_v \gets \text{std}(H) / \text{mean}(H)$\;
\uIf{$c_v < \eta$ \KwAnd $|L_B| > \epsilon$}{
    $S' \gets [B.pkts$ \KwFor $B$ \KwForIn $L_B]$\;
    $S.\text{union}(S')$;
}
}
\KwRet $S$;
\caption{Key Packet Extraction}
\label{algo:key_packet}
\end{algorithm}

\subsection{Neighboring Key Packet Distribution and Decision Tree} 
DeviceRadar uses the packet neighboring probability matrix and the set of key packets to form a feature vector for device identification. To reduce the complexity of the feature vector, we only use the first $N$ key packets sorted by their periods in ascending order, as more frequent key packets are more helpful for timely identification. Suppose a series of packets from one IP address are observed during a time window $T_w$, and $\boldsymbol{v}$ is the feature vector of the neighboring key packet distribution. Given the key packets $[x_1, x_2, ..., x_N]$, the generation of such a feature vector consists of the following steps: 
\begin{enumerate}[label=\roman*.,topsep=0pt,itemsep=0pt,parsep=0pt,partopsep=0pt]

\item Initialize a vector $\boldsymbol{v}$ of $N$ zeros.

\item For each packet $p_i$, obtain the probability $\mathrm{P}(\mathcal{X}=x_j|p_i;\mathbb{D})$ from the matrix $M_\mathbb{D}$.

\item Add the probability to the corresponding position of the vector, i.e., $\boldsymbol{v}[j] = \boldsymbol{v}[j] + \mathrm{P}(\mathcal{X}=x_j|p_i;\mathbb{D})$.

\item Iterate step {\romannumeral3} for all the $N$ key packets.

\item Iterate step {\romannumeral2} and step {\romannumeral3} for all the observed packets during the time window.
\end{enumerate}

This feature vector meets the typical input of simple ML models like tree classifiers~\cite{tabular}.
Furthermore, it includes the implicit semantics learned from the embedding space, which enable simple ML models to sense spatial relationships even though these models may not naturally have this ability. We choose the decision tree of CART (classification and regression tree) as the final classifier because of its good deployability on the data plane. A per-device classifier is trained for each target device. Each training sample on a node $\mathcal{N}$ can be denoted by $(\boldsymbol{v}, y)$, where $y \in \{0, 1\}$ is the label for the presence of the device. The impurity $I$ of $\mathcal{N}$ is calculated by the proportion of labels in $\mathcal{N}$, e.g., using Gini impurity:
\begin{equation}
\begin{aligned}
    I = 1 - \sum_{i \in \{0, 1\}}\mathrm{P}_{i}^2,~\text{where}~\mathrm{P}_i = \frac{1}{|\mathcal{N}|} \sum_{(\boldsymbol{v}, y) \in \mathcal{N}} \mathbb{I}\{y = i\}.
\end{aligned}
\end{equation}
The training process iterates to maximize the impurity decrease by finding a decision that splits the node into two nodes. After the training and during the inference, a sample $(\boldsymbol{v}, y)$ will go through decision paths and fall into a leaf node $\mathcal{T}$, and the predicted label is determined by:
\begin{equation}
    \hat{y} = \arg\max\limits_{y} \frac{1}{|\mathcal{T}|} \sum_{(\boldsymbol{v}, y) \in \mathcal{T}} \mathbb{I}\{y = i\}.
\end{equation}

\textcolor{black}{
\subsection{Analysis of Computational Complexity}
}

\textcolor{black}{
We theoretically analyze the computational complexity of each step in DeviceRadar, as outlined in Table~\ref{tab:complexity}. It can be seen that the computational complexity of DeviceRadar is proportional to the target device packet number $|P_\mathbb{D}|$, the number of extracted key packets $N$, and the number of feature vectors $|V|$ as training data. Importantly, DeviceRadar does not involve operations with complexities exceeding quadratic terms concerning traffic volume. This result supports the practical implementation of DeviceRadar in high-speed networks.
}

\begin{table}[t]
    \centering
    \caption{\textcolor{black}{Complexity of DeviceRadar; $V$ is the set of feature vectors from each time window $T_w$ for training decision trees.}}
    \begin{tabular}{c|c}
    \toprule
    \textcolor{black}{Component} & \textcolor{black}{Complexity} \\
    \midrule
    \textcolor{black}{Packet Embedding} & \textcolor{black}{$O(2\cdot c \cdot k \cdot |P_\mathbb{D}|)$} \\
    \textcolor{black}{Packet Neighboring Probability Matrix} & \textcolor{black}{$O(1)$} \\
    \textcolor{black}{Key Packet Extraction} & \textcolor{black}{$O(|P_\mathbb{D}|)$} \\
    \textcolor{black}{Neighboring Key Packet Distribution} & \textcolor{black}{$O(N)$} \\
    \textcolor{black}{Decision Tree Classifier} & \textcolor{black}{\makecell[l]{Training: $O(N \cdot |V|\log |V|)$ \\ Inference: $O(tree\_depth)$}} \\
    \bottomrule
    \end{tabular}
    \label{tab:complexity}
\end{table}

\section{DeviceRadar Data Plane}
In this section, we elaborate on the transformation and deployment of the device fingerprinting components from the control plane to the data plane. Particularly, the components are transformed into programmable switch-compliant data structures for online use on the data plane.

\subsection{Metadata in P4 Program}
To deploy DeviceRadar on the data plane, we write a P4 program to implement a complete suite of components for device fingerprinting generated on the control plane, as described in Section~\ref{sec:control_plane}.
These will be used at runtime, including 1) the acquisition of directional packet sizes;
2) the packet neighboring probability matrix;
3) the maintenance of a feature vector;
and 4) the decision tree.

A PISA programmable switch parses no protocols unless they are explicitly defined by the installed P4 program. In our program, we parse the IPv4 header with three fields:

\begin{itemize}[topsep=0pt,itemsep=0pt,parsep=0pt,partopsep=0pt]
\item Source address: \texttt{ip4Addr\_t srcAddr}

\item Desination address: \texttt{ip4Addr\_t dstAddr}

\item IP packet size: \texttt{bit<16> totalLen}
\end{itemize}

The programmable switch offers some intrinsic metadata about the switch. We mainly use the ingress port field \texttt{ingress\_port}, which is either LAN or WAN, to decide the packet direction.


Besides, a P4 program allows a packet to carry some user-defined metadata through the match-action pipeline. We define the following metadata in our P4 program:

\begin{itemize}[topsep=0pt,itemsep=0pt,parsep=0pt,partopsep=0pt]
\item Directional packet size: \texttt{bit<16> dir\_size}

\item Neighboring packet probability: \texttt{bit<32> prob\_x}; $x=1, 2, ..., N$ for each key packet

\item Value of a feature vector: \texttt{bit<32> v\_x}; $x=1, 2, ..., N$ for each dimension of a feature vector

\item Packet timestamp: \texttt{bit<32> tstamp}

\item Timeout sign of a time window: \texttt{bit<8> timeout} 

\item Device identification result: \texttt{bit<32> label}
\end{itemize}

\subsection{The Match-Action Pipeline}
\label{subsec:mat}
A packet will go through the match-action pipeline defined by the P4 program. We design four match-action tables, which transplant the complete function of models constructed on the control plane to the data plane, and bypass the operation and resource constraint of programmable switches. 
In particular, we manage to handle the unsupported floating-point numbers of the probability matrix by P4-compliant data types. Moreover, our match-action tables utilize the P4 registers to realize the incremental construction of feature vectors, which guarantees the high throughput of processing at line rate.

\textbf{Direction Table} (Listing~\ref{list:direction_table}). This table achieves the acquisition of directional packet sizes by identifying the ingress port and applying the logic of Equation~(\ref{eq:dir_ps}). Only two rules for the two directions will be written in this table.

\begin{lstlisting}[caption=Sample P4 code of direction table., label=list:direction_table, language=p4, mathescape]
action set_lan_packet_size() {
    meta.dir_size = hdr.ipv4.totalLen + 1500;
}
action set_wan_packet_size() {
    meta.dir_size = hdr.ipv4.totalLen;
}
table directional_packet_size {
    key = {ig_intr_md.ingress_port: exact;}
    actions = {
        set_lan_packet_size;
        set_wan_packet_size;
    }
}
\end{lstlisting}

\textbf{Probability Table} (Listing~\ref{list:prob_table}). This table realizes the packet neighboring probability matrix. Given a directional packet size, it adds the probability of neighboring packets being the $N$ key packets into the metadata. For a target device $\mathbb{D}$, the number of rules written in the table equals the total number of directional packet sizes in the traffic $P_\mathbb{D}$, and the number of parameters for the action equals $N$.
However, the item of the probability matrix $M_{\mathbb{D}}$ is a floating-point number in the range of $[0, 1]$, which is not supported in P4. To resolve this problem, we scale the values in $M_{\mathbb{D}}$ to the range of $[0, 255]$ and round down to integers, which can be assigned to the first 8 bits of the P4-compliant data type \texttt{bit<32>}.

\textit{Analysis of accuracy loss:} The numerical space after scaling and rounding has 256 distinct values. The accuracy loss of mapping the original numerical space (real number between 0 and 1, suppose uniformly distributed) to this space can be approximated by $ (1 - 0) / 256 = 3.906 \times 10^{-3}$, which is trivial compared to the original numerical space. 

\begin{lstlisting}[caption=Sample P4 code of probability table., label=list:prob_table, language=p4, mathescape]
action set_meta_prob(bit<32> prob_1, ...) {
    meta.prob_1 = prob_1;
    ...
}
table packet_size_to_prob
{
    key = {meta.dir_size: exact;}
    actions = {
        set_meta_prob;
    }
}
\end{lstlisting}

\textbf{Stateful Update Table} (Listing~\ref{list:stateful_table}). 
A feature vector is obtained by observing and aggregating the packets within a time window. However, in P4 switch ASICs, the parsed header and metadata inside a packet are immediately re-instantiated when the packet is sent out of the switch and cannot be reaccessed. To deal with it, one approach is to store the parsed header and metadata for every packet, which is obviously resource-consuming. Hence, we realize an approach to the incremental construction of feature vectors using P4 registers that support stateful storage. In line 1 of the sample code, a 3D register array is declared. It can maintain an $N$-dimensional feature vector for each of \texttt{IP\_SIZE} IP addresses at most. The action of the table is to add each of the $N$ neighboring probabilities, obtained by the previous table and stored in the metadata, to the corresponding positions of registers, and store the updated feature vector in the metadata. Once a window $T_w$ is timed out, which can be identified by the timestamp metadata \texttt{tstamp}, the last packet carries the feature vector of this window in the metadata \texttt{v\_x} and sets the metadata \texttt{timeout} to 1. This \texttt{timeout} sign will trigger the inference table for the final result of device identification. 

A register may encounter overflow with the accumulation of metadata before being emptied at the end of a time window. Given that \texttt{prob\_x} occupies 8 bits, in an extreme case that all packets are from one IP address, a 32-bit register can support the accumulation of at least $2^{32}/2^8 = 2^{24} 	\approx 16$ million packets without overflow. According to the statistics~\cite{caida}, the average packet rate in a backbone network is about 600 Kpps. It means that a register is guaranteed to be safe even if the time window is set to tens of seconds.

\begin{lstlisting}[caption=Sample P4 code of stateful update table., label=list:stateful_table, language=p4, mathescape]
Register<bit<32>, bit<N>>(IP_SIZE) reg1;
RegisterAction<bit<32>, bit<N>, bit<32>>(reg1)
reg1_prob_update = {
    void apply(inout bit<32> reg_data, out bit<32> rtn) {
        reg_data = reg_data + meta.prob_1;
        rtn = reg_data;
    }
};
action action_update_register1() {
    meta.v_1 = reg1_prob_update.execute(0);
}
table stateful_update_1 {
    actions = {
        action_update_register1;
    }
}
\end{lstlisting}

\textbf{Inference Table.} A sample of the inference code is shown in Listing~\ref{list:inf_table}. 
This table implements the final model, i.e., the decision tree, for device identification. 
For a decision tree trained on the control plane, we write a script that extracts each path from the root to a leaf node as a decision rule, in which the label on the leaf can be determined by the range of certain features. Thus, we design the inference table on the data plane using the P4 range match for each dimension of feature vectors and using the P4 exact match for the timeout sign. 
We also use the pruning technique to empirically set the maximum number of leaf nodes to 500, which guarantees that the rules fit in a single stage and also prevents overfitting. The inference result is stored in the metadata and can be used as prior knowledge by subsequent actions, such as redirecting, throttling or filtering the traffic from the high-risk devices.

\begin{lstlisting}[caption=Sample P4 code of inference table., label=list:inf_table, language=p4, mathescape]
action set_label(bit<8> label) {
    meta.label = label
}
table node {
    key = {
        meta.timeout: exact;
        meta.v_1: range;
        meta.v_2: range;
        ...
        meta.v_N: range;
    }
    actions = {set_label;}
}
\end{lstlisting}

\section{Evaluation}
We evaluate DeviceRadar by answering the following questions:

\noindent1) Can DeviceRadar accurately identify target devices within traffic that middleboxes have modified? (Section~\ref{sec:eval1})

\noindent2) Can DeviceRadar achieve real-time processing and high throughput in high-speed networks? (Section~\ref{sec:eval2})

\noindent3) Is there any use case to highlight the advantage of DeviceRadar as a part of a defense system? (Section~\ref{sec:eval3})

\subsection{Implementation and Testbed}
We prototype the complete framework for evaluation. The control plane components are mainly implemented by Python 3.6 with over 500 lines of code, and the data plane components are implemented by $\mathrm{P}4_{16}$ with over 1300 lines of code.

\textbf{Control Plane.} The controller is a general computing server with two CPUs (Intel(R) Xeon(R) Gold 5117, 28 cores), two GPUs (NVIDIA GeForce RTX 2080 SUPER GPUs, 8GB), 128GB memory and Ubuntu 18.04.1 (Linux 5.4.0-80-generic). The packet embedding is implemented by PyTorch 1.10.1, and the decision tree is implemented by scikit-learn 0.24.1. The server is installed with Intel P4 Studio to control the switch.

\textbf{Data Plane.} We use a Wedge100BF-65X data center switch with a programmable Tofino programmable switch ASIC~\cite{tofino}, supporting 4 $\times$ 10 GbE switching via breakout cables. To evaluate the processing speed and throughput of DeviceRadar, we employ a commercial network tester (Keysight XGS12) to generate high-speed traffic at a given rate. This setting simulates the realistic switching rates in an ISP network.

\textbf{IoT Testbed.} We configure a real-world IoT testbed for traffic collection, including 14 types of off-the-shelf devices that cover most mainstream IoT manufacturers in China (e.g., Huawei, Xiaomi, Skyworth) and popular types of devices (e.g., camera, plug, sound box). The devices are placed in an open laboratory where staff (usually 4$\sim$6 people) are free to use the devices\footnote{Staff were informed about the experiment; we only capture traffic data but do not save any privacy-related data, such as video recordings.}. 
To collect the traffic that can be actually seen by ISPs (i.e., traffic modified by middleboxes), we use an enterprise router to function as middleboxes like NAT and VPN, and employ a personal computer connected to the router as the traffic collector. Specifically, we use NAPT and OpenVPN in our evaluation. More information about the IoT devices and the testbed is available in the Appendix.

\subsection{Experimental Setup}
\textbf{Baselines.} To measure the improvement of DeviceRadar, we establish three types of methods as baselines:

\noindent 1) \textit{Signature-based method}: We use the state-of-the-art approach, PingPong~\cite{pingpong}, that considers NAT and VPN scenarios. It extracts the predicted sizes of TCP packet pairs (i.e., device to cloud and cloud to device) as signatures to detect IoT devices. Some other approaches inherently not for online and real-time use (e.g., based on DNS queries) are excluded.

\noindent 2) \textit{ML-based method}: For those using traffic statistics, we utilize DarkSide~\cite{darkside} that calculates 16 statistics for packet sizes and inter-arrival time in a window of 180 packets, and uses a Random Forest classifier for device identification. For those using header fields, we employ DeNAT~\cite{denat}, which parses 9 header fields by NetFlow such as network portion of IP, port, protocol, timestamp, TCP flag and ToS. The LGBM algorithm is selected for the best accuracy in its evaluation.

\noindent 3) \textit{DL-based method}: We use HomeMole~\cite{secret}, a state-of-the-art IoT device identification approach from the view of ISPs. It leverages a deep learning model of bidirectional LSTM to learn the temporal relations in every 100 packets. It claims to be effective in complex networks including NATs and VPNs.

All the baseline methods are reproduced according to their published paper or released code, except for PingPong, which has a small modification. Given that PingPong is to identify an IoT event (e.g., light bulb turning ON/OFF) when the event occurs, and both our dataset and public datasets do not precisely record the time of the events, we use a sliding time window of 10 seconds to split the consecutive packets into sequences as traffic samples for device identification.

\textbf{Datasets.} We use four IoT traffic datasets and one background traffic dataset in total.

\textit{Self-built IoT dataset}: Collected by our own IoT testbed with 14 distinct devices. Specifically, it is composed of three 10-day traffic datasets, collected in March, June and September in 2021, each of which contains about 10 GB PCAP data. In the rest of the paper, ``Self-built'' 
refers to the dataset of March, and ``March''/``June''/``September'' indicates the three datasets in specific experiments.

We further use three public IoT traffic datasets as benchmarks. Their devices are purchased and placed in the U.S., U.K. and Australia, respectively. These datasets supplement the IoT manufacturers and cloud providers of IoT devices in different countries to improve the generality of the experiments. Note that some of the devices are not used in our experiments either because their traffic data are somehow empty or contain very few packets or we find they are so frequently disconnected that the packets for device initialization, such as DHCP, DNS and NTP, dominate their traffic. To summarize, the public datasets are as follows:

\noindent1) \textit{NEU}~\cite{iot_measure}: Collected by Northeastern University (U.S.), containing 1.8 GB PCAP files of traffic from 26 devices collected in 3 days.

\noindent2) \textit{ICL} ~\cite{iot_measure}: Jointly collected by Imperial College London (U.K.) and NEU, containing 488 MB PCAP files of traffic from 22 devices collected in 3 days.

\noindent3) \textit{UNSW}~\cite{unsw}: Collected by the University of New South Wales (Australia), containing 2.7 GB of PCAP files of traffic from 15 devices collected in 10 days.

\textcolor{black}{
As a result, our experiments have covered 77 IoT devices from 4 regions in total, including cameras, speakers, hubs, plugs, bulbs, thermostats, and even microwaves and fridges.
The full list of devices is available in the Appendix.
}

\begin{figure}[htbp]
\centering
\subfigure[Packet size distribution]{
\begin{minipage}{0.46\linewidth}
\centering
\includegraphics[width=\textwidth]{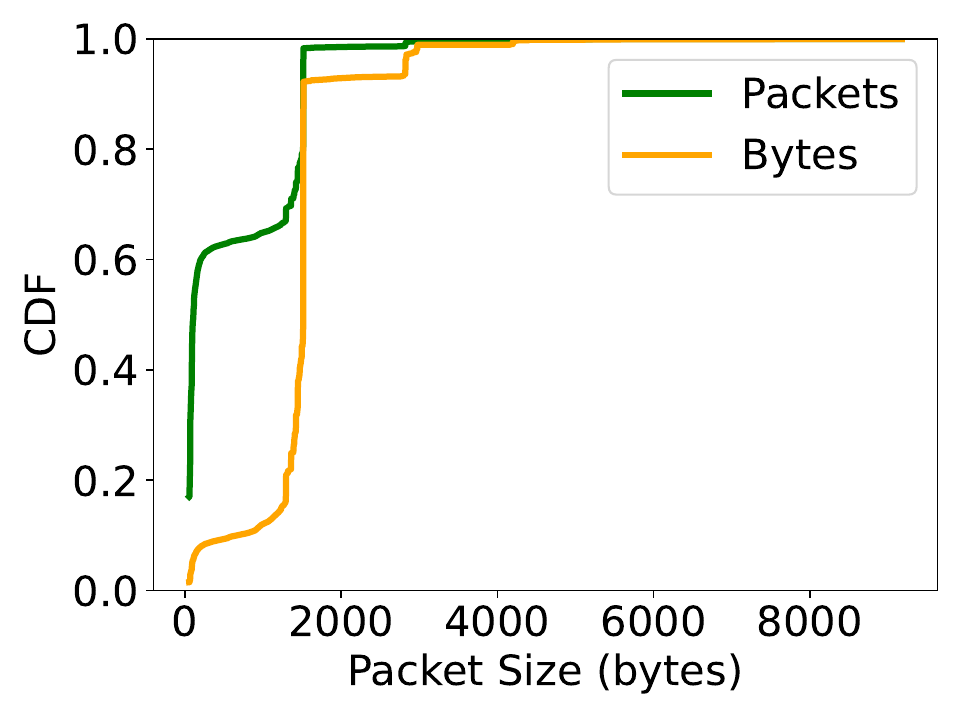}
\end{minipage}
}
\subfigure[Protocol breakdown]{
\begin{minipage}{0.46\linewidth}
\centering
\includegraphics[width=\textwidth]{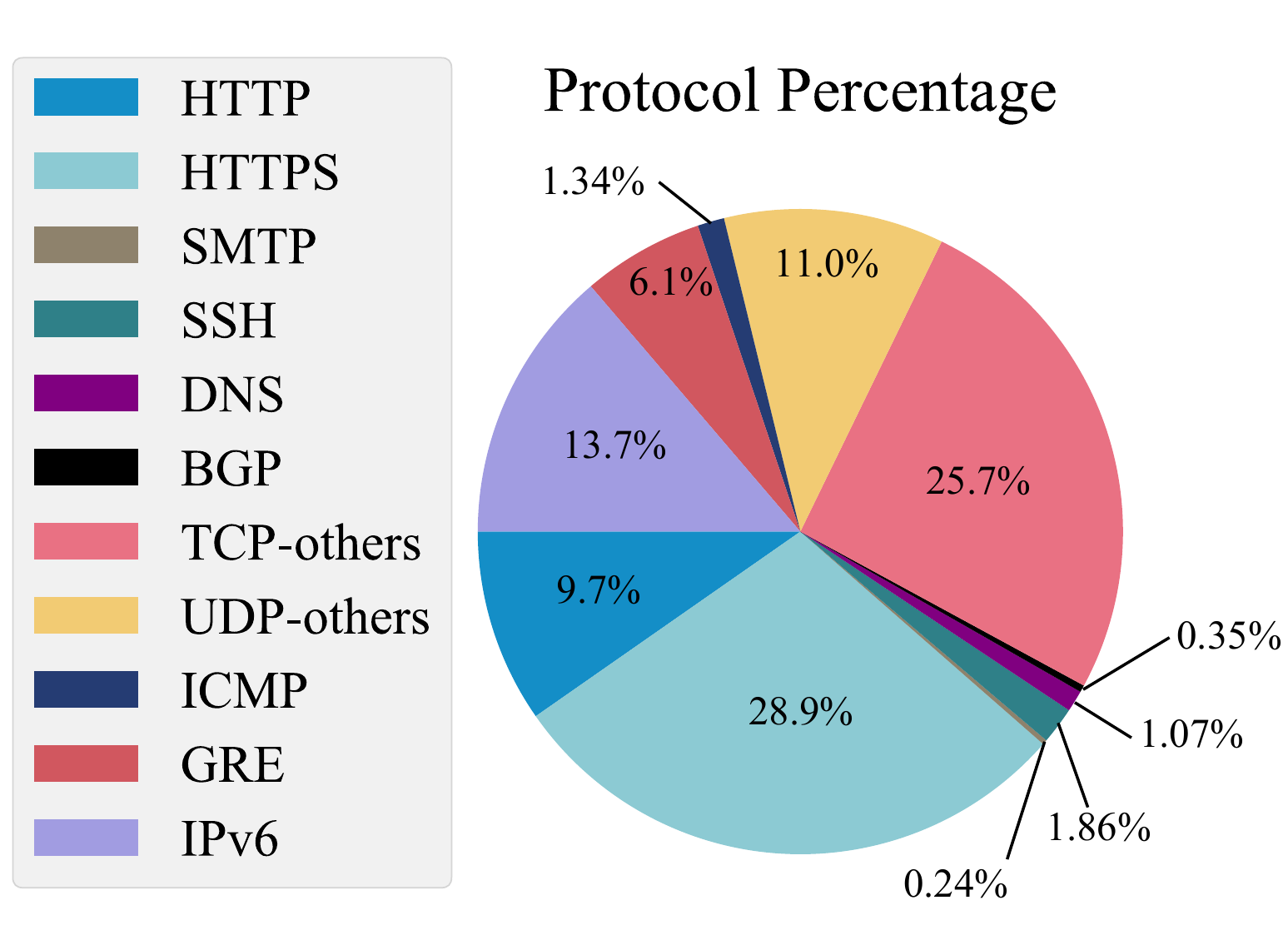}
\end{minipage}
}
\caption{Real-world ISP background traffic statistics.}
\label{fig:mawi}
\end{figure}

To reflect the diversity of real-world traffic for ISPs, we add a public dataset as the background traffic~\cite{mawi}, which is collected by the MAWI Working Group at the transit link of WIDE backbone to the upstream ISP. The WIDE project offers a series of 15-minute real-world traces in a continuous period of time, and each of them occupies 3.35 GB to 17.07 GB of PCAP files. The average rate ranges from 175.76 Mbps to 1099.08 Mbps at different time. We use the dataset of April in 2022 that demonstrates an up-to-date view of networks. The statistics of packet size distribution and protocol breakdown are illustrated in Fig.~\ref{fig:mawi}, qualifying the diversity of the traffic.

\textbf{Trace Filter and Mixture.} For IoT traffic datasets, we filter the traffic that an ISP will not see, including DHCP, ARP, DNS resolved by local servers, local broadcasting traffic like SSDP, and the traffic between devices in the same LAN. For background traffic, we retain all the packets. 

An issue in some of the trace mixture settings (e.g., in~\cite{pinpointing}) is that, though the IoT traffic of various devices is fused by a middlebox (e.g., with a uniform source IP by NAT), the background traffic is not well mixed and can be easily stripped out by its original network artifacts, such as totally different IP addresses. It can be interpreted as a situation where only IoT traffic goes through a middlebox but is not aggregated with other background traffic by the next-hop device, which can also be a middlebox. Without loss of generality, we consider a more challenging setting by assuming all traffic from a specific link is modified by a middlebox. We replay both the IoT traffic and background traffic simultaneously through the middlebox in our testbed (by NAT or VPN) and use the modified mixed traffic for the subsequent experiments.

\textbf{Packet Labelling.} 
We refer to the method in~\cite{secret} to obtain the ground truth labels of the modified packets after the replay (i.e., which device it belongs to). We simultaneously collect the traffic before and after the middlebox, and label a modified packet by finding another packet before the middlebox that satisfies three conditions: 1) they have the same direction; 
2) the timestamp difference of two packets is less than 0.02 seconds; 
and 
3) in the VPN scenario, the length of the packet before the middlebox is slightly smaller than the one after the middlebox due to encryption. We manually check 1\% of random samples of the packets labeled by this method and the correctness rate is over 98\%.

\begin{figure*}[t]
    \centering
    \subfigure[Self-built (NAT)]{
    \begin{minipage}{0.23\textwidth}
    \centering
    \includegraphics[width=\textwidth]{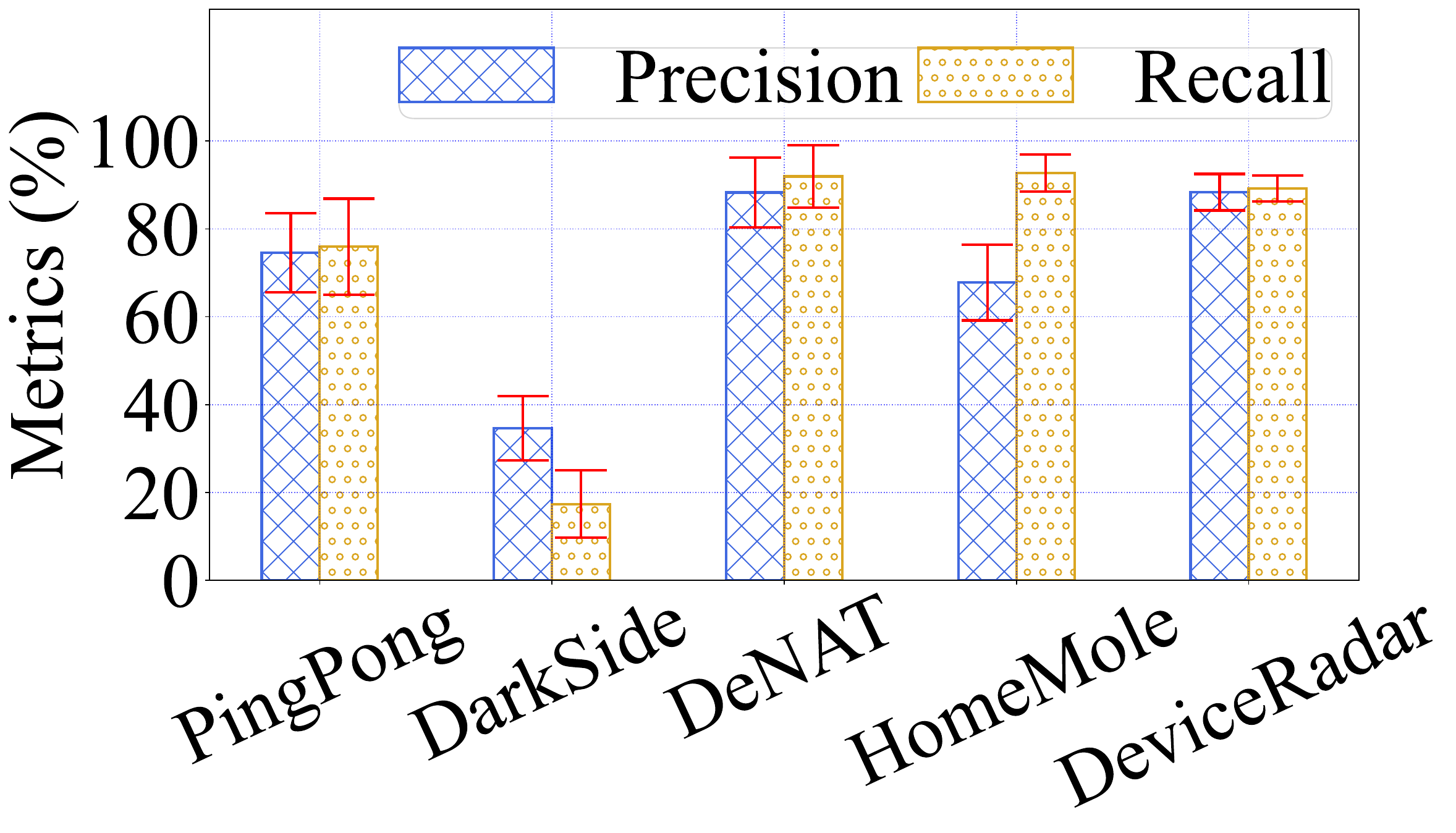}
    \end{minipage}
    }
    \subfigure[NEU (NAT)]{
    \begin{minipage}{0.23\textwidth}
    \centering
    \includegraphics[width=\textwidth]{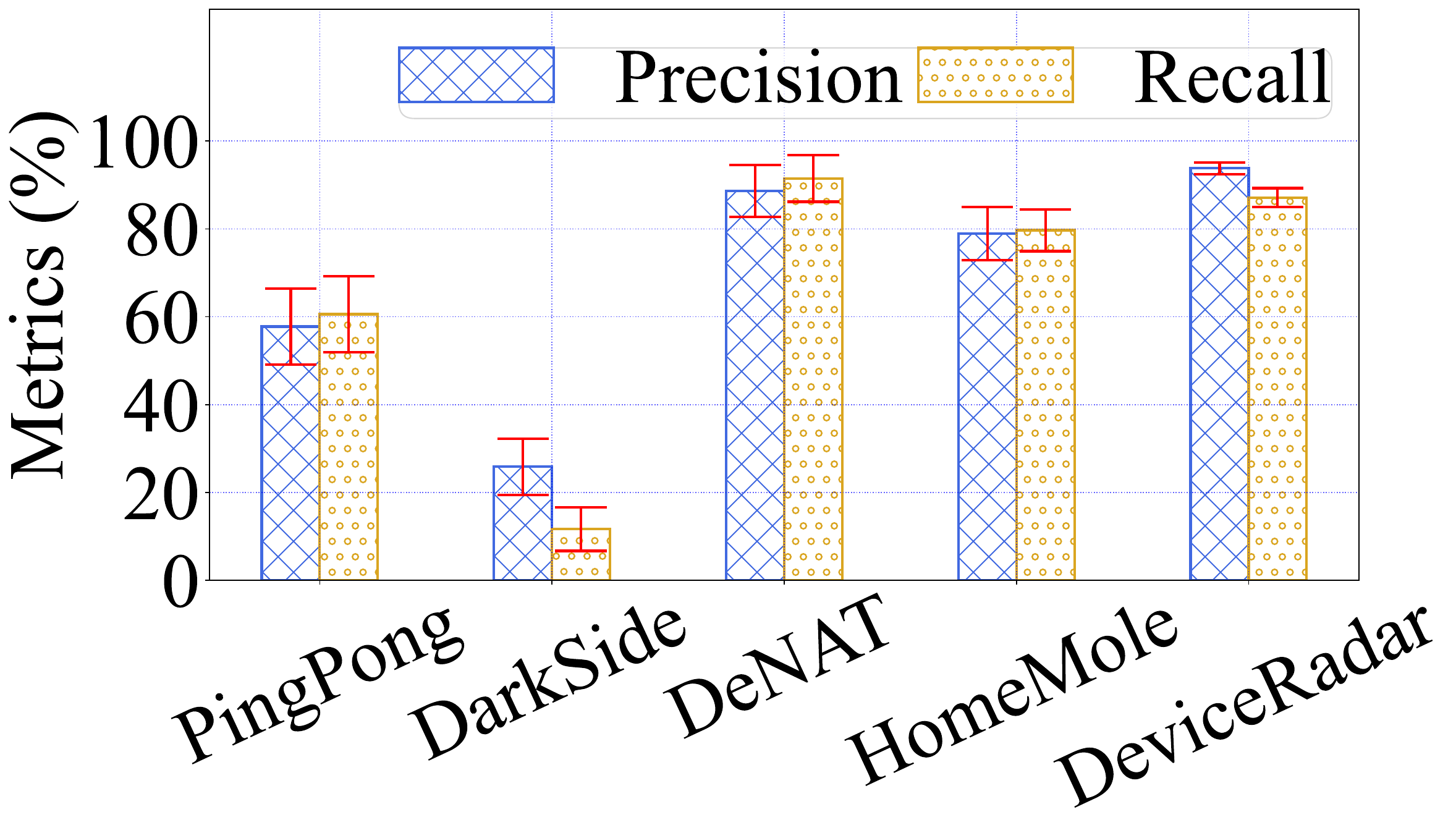}
    \end{minipage}
    }
    \subfigure[ICL (NAT)]{
    \begin{minipage}{0.23\textwidth}
    \centering
    \includegraphics[width=\textwidth]{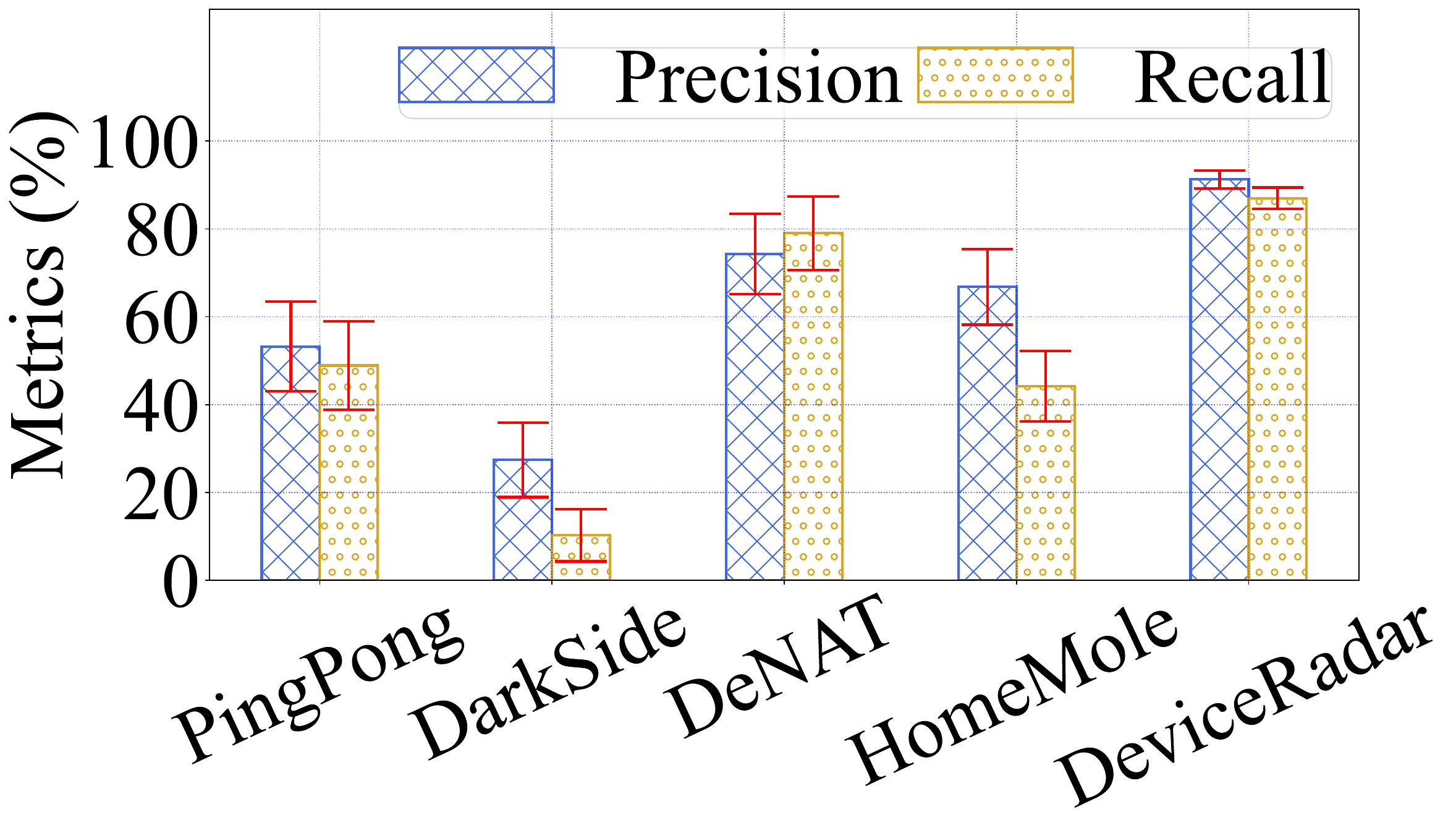}
    \end{minipage}
    }
    \subfigure[UNSW (NAT)]{
    \begin{minipage}{0.23\textwidth}
    \centering
    \includegraphics[width=\textwidth]{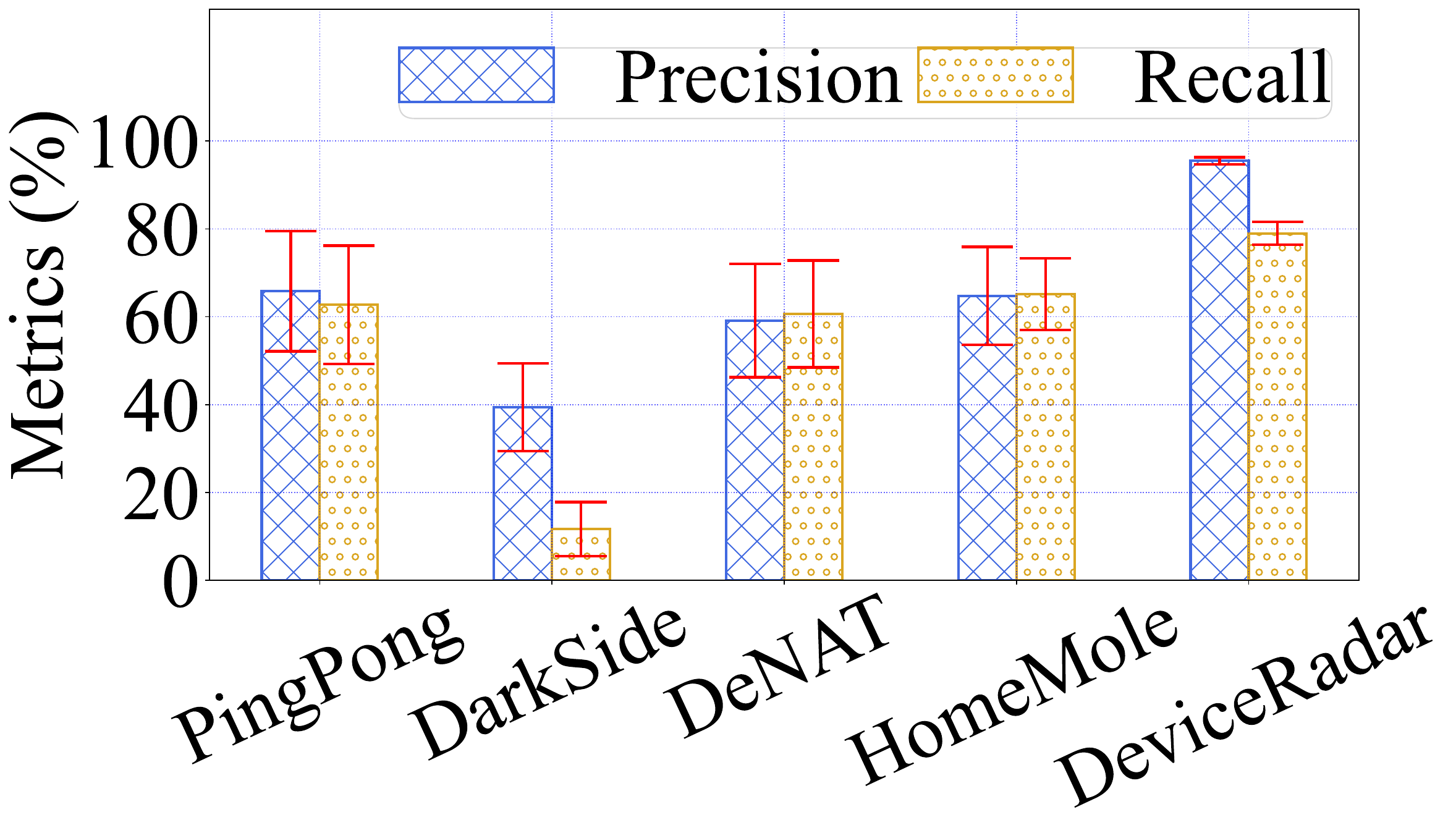}
    \end{minipage}
    }
    \subfigure[Self-built (VPN)]{
    \begin{minipage}{0.23\textwidth}
    \centering
    \includegraphics[width=\textwidth]{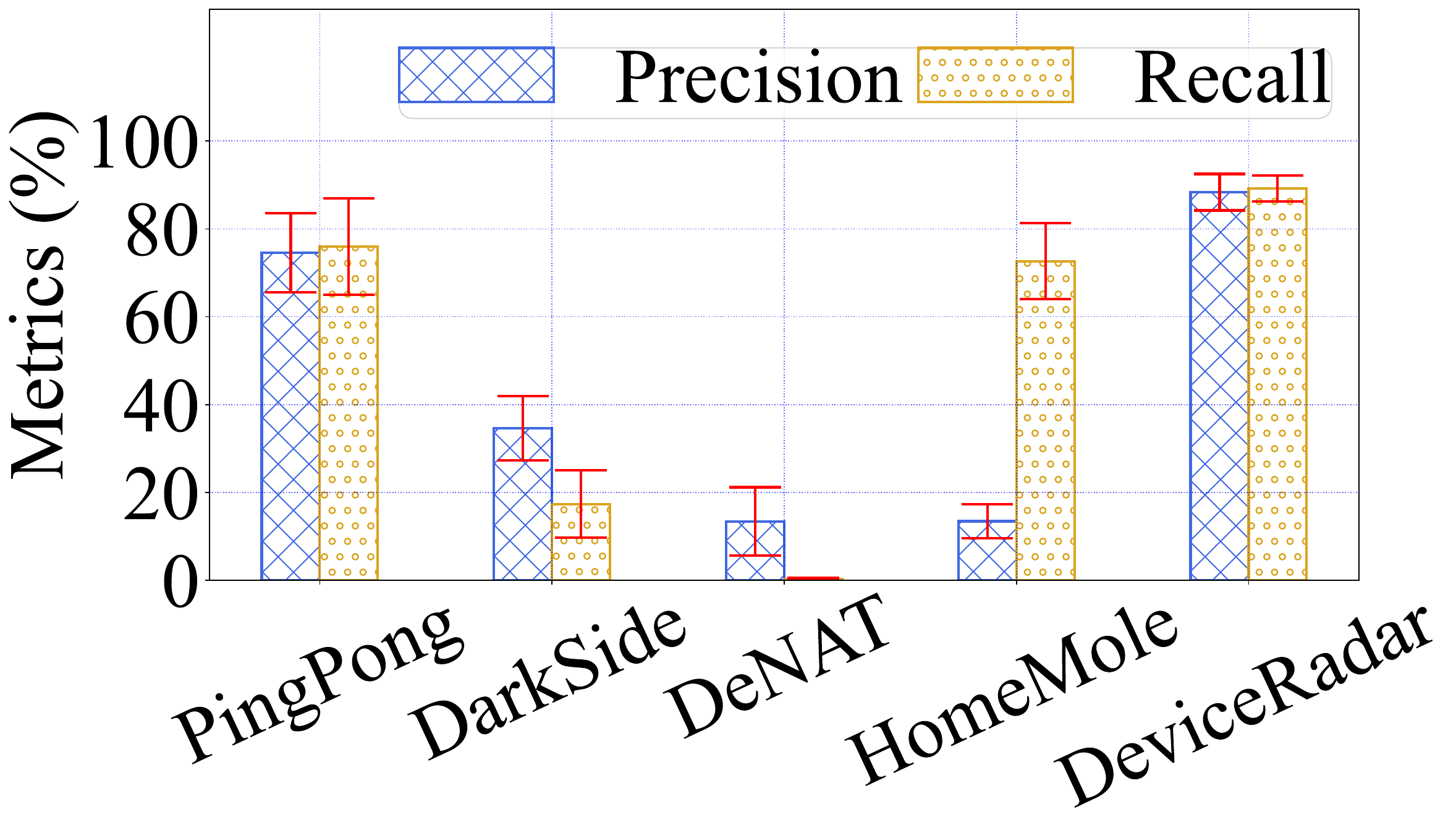}
    \end{minipage}
    }
    \subfigure[NEU (VPN)]{
    \begin{minipage}{0.23\textwidth}
    \centering
    \includegraphics[width=\textwidth]{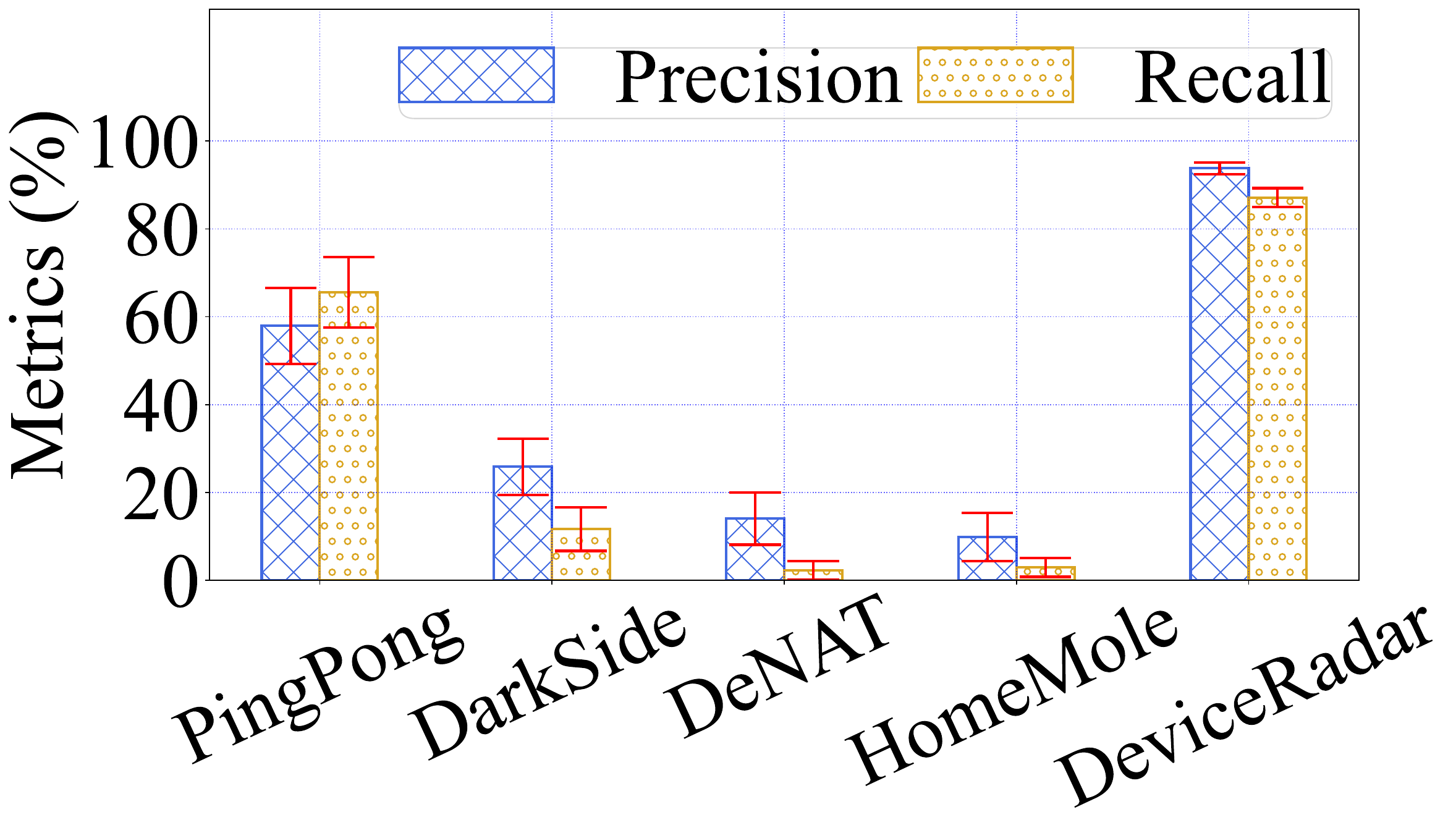}
    \end{minipage}
    }
    \subfigure[ICL (VPN)]{
    \begin{minipage}{0.23\textwidth}
    \centering
    \includegraphics[width=\textwidth]{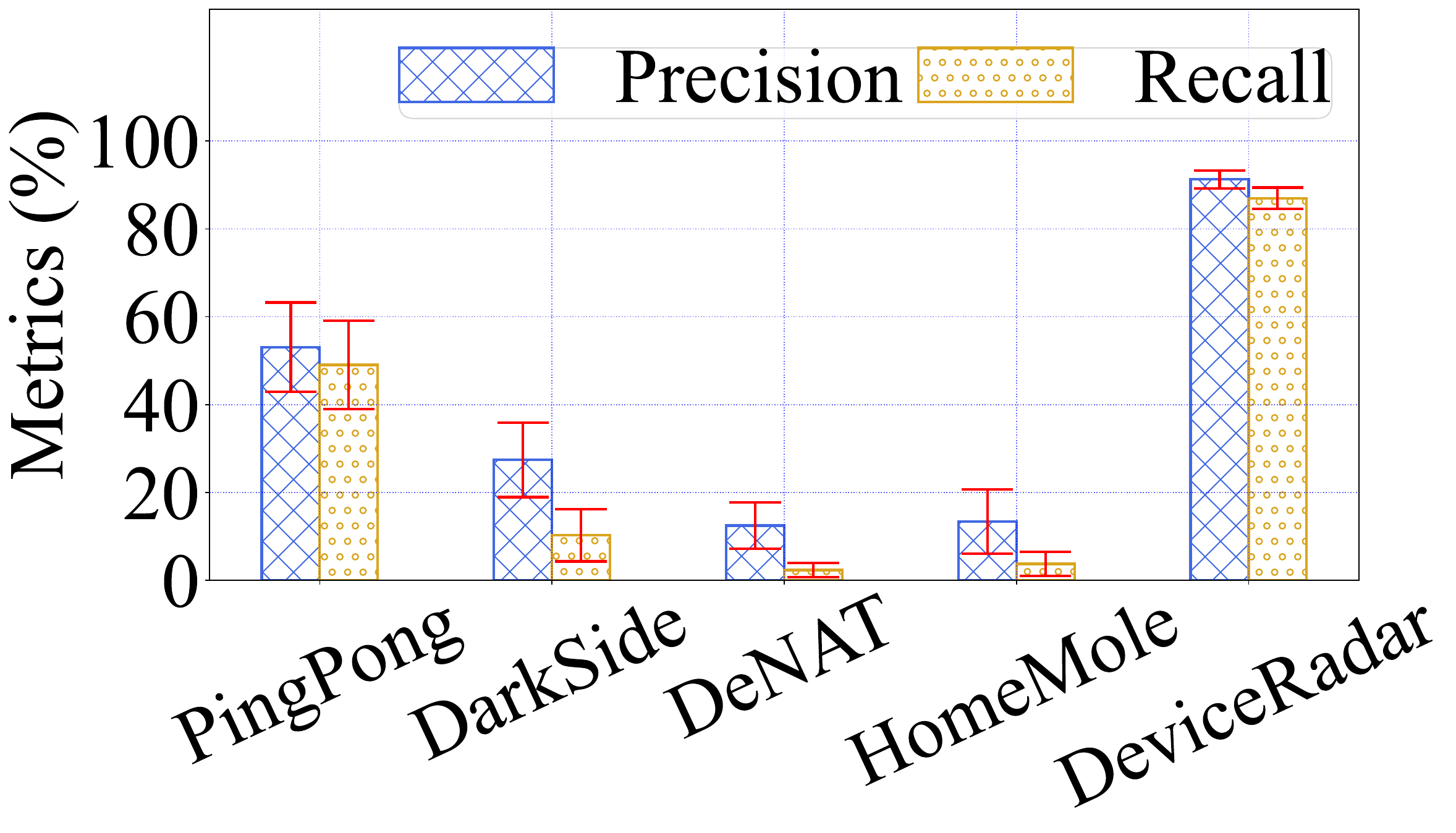}
    \end{minipage}
    }
    \subfigure[UNSW (VPN)]{
    \begin{minipage}{0.23\textwidth}
    \centering
    \includegraphics[width=\textwidth]{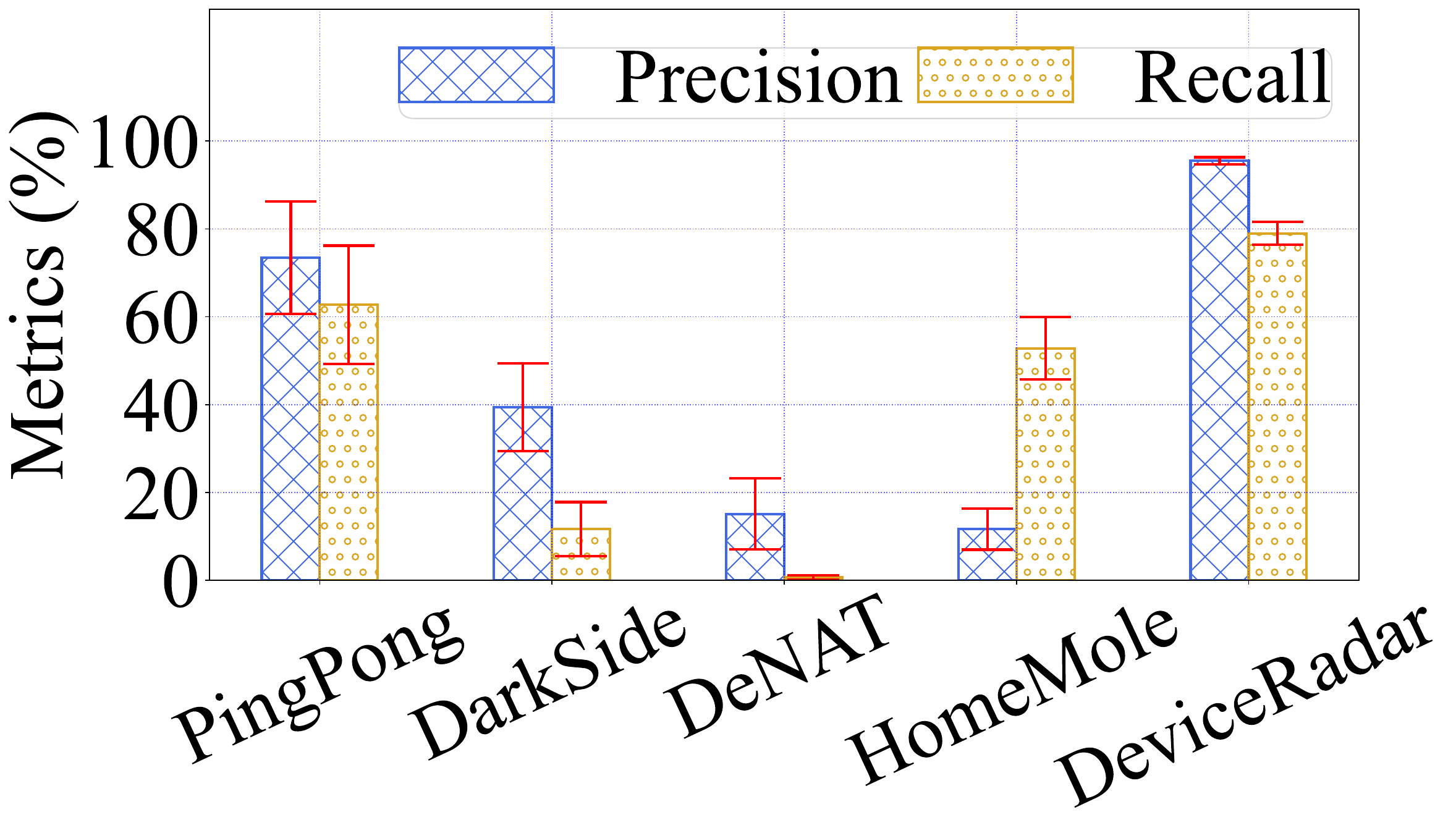}
    \end{minipage}
    }
    \caption{Device identification accuracy comparison (bar -- average metric value; red line -- standard error).}
    \label{fig:accuracy}
\end{figure*}

\begin{figure}[t]
\centering
\subfigure[Precision \& Recall]{
\begin{minipage}{0.46\linewidth}
\centering
\includegraphics[width=\textwidth]{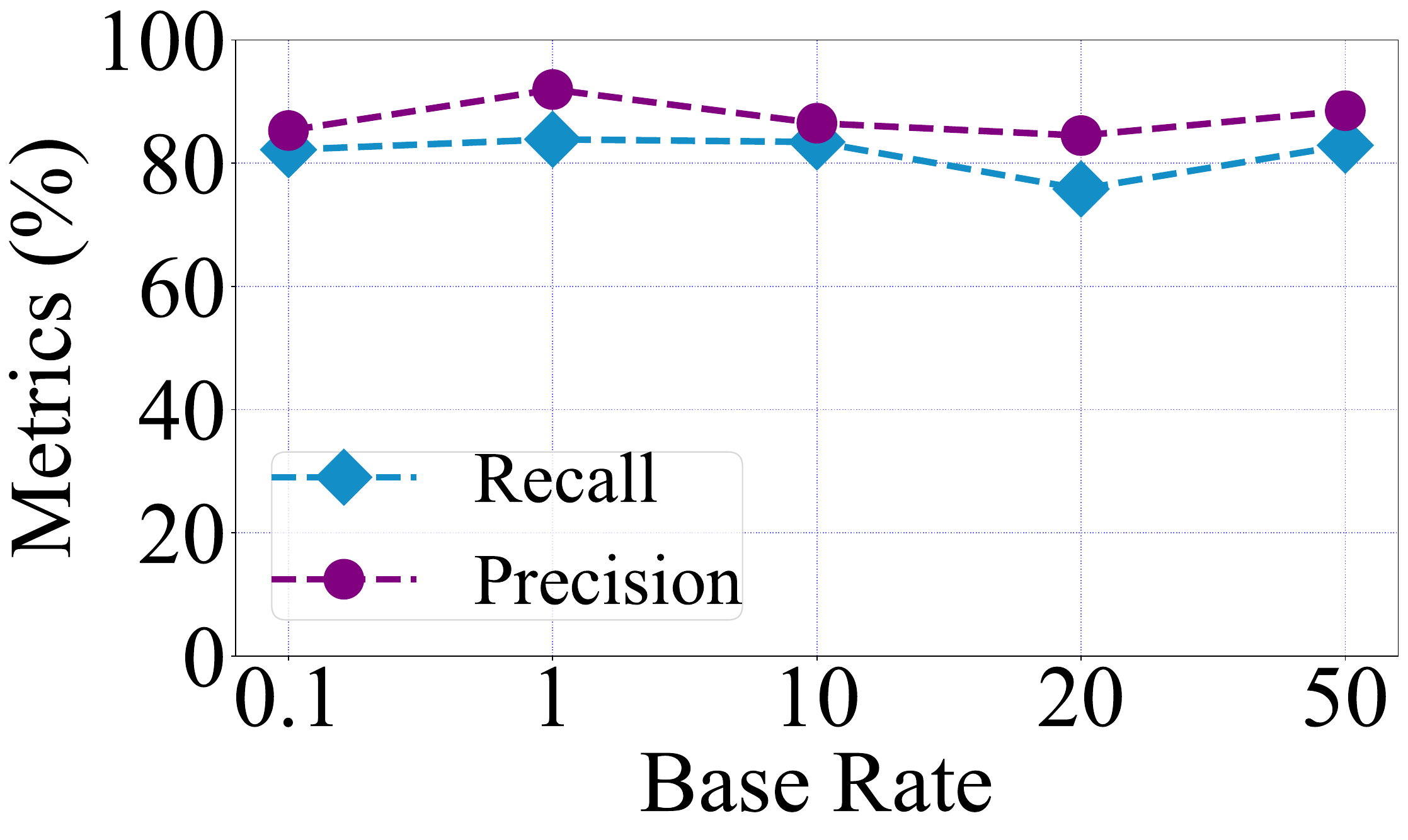}
\end{minipage}
}
\subfigure[False Positive Rate]{
\begin{minipage}{0.46\linewidth}
\centering
\includegraphics[width=\textwidth]{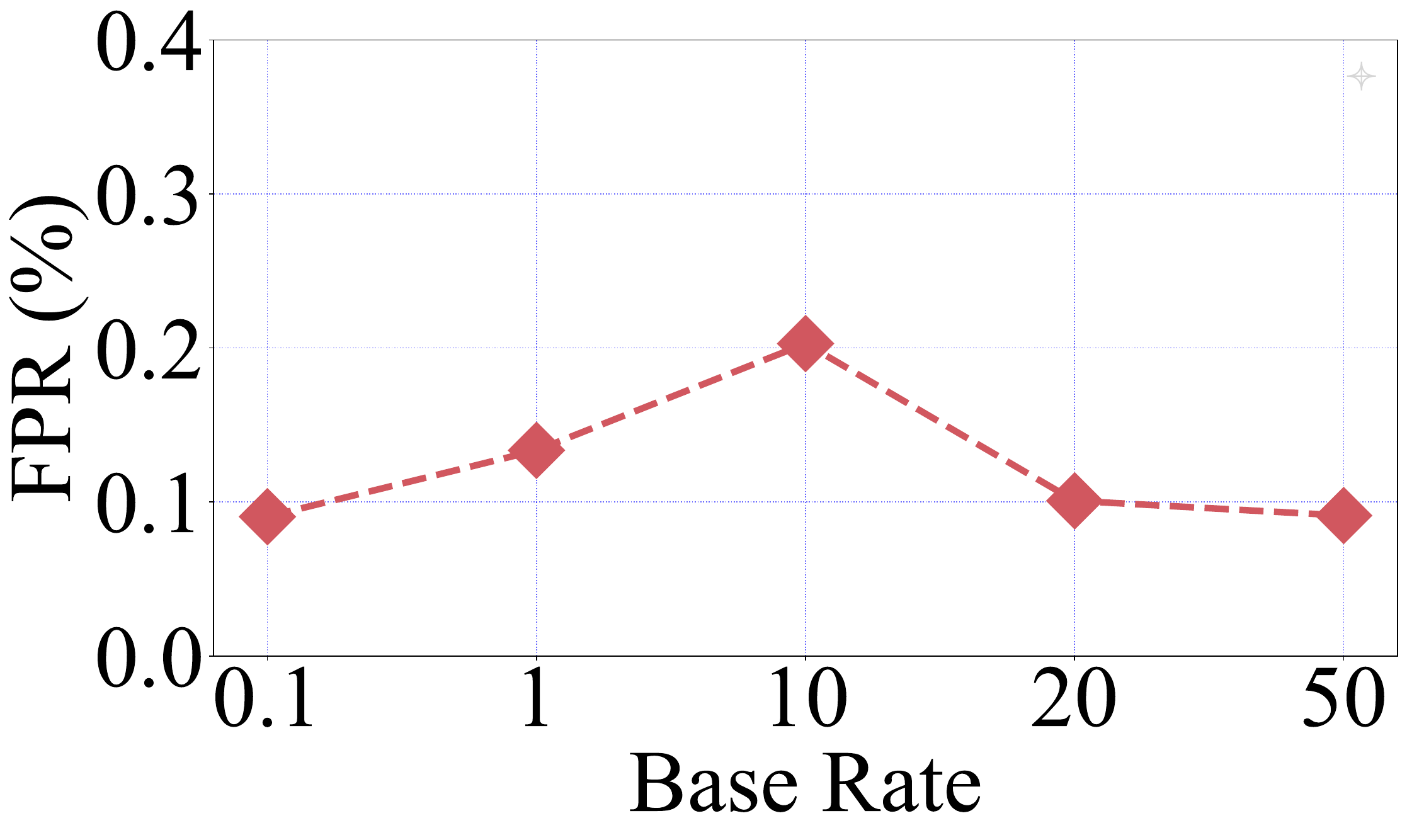}
\end{minipage}
}
\caption{Base rate versus precision, recall and FPR.}
\label{fig:base-rate}
\end{figure}

\textbf{Data Splitting and Labelling.} We follow the data splitting strategy of each method to generate data samples (i.e., feature vectors) for training and testing, including host-level packet window for DarkSide, flow-level packet window for HomeMole, Netflow records for DeNAT, and time window for PingPong and DeviceRadar. 
We empirically set the time window of DeviceRadar to 1 second and discuss its impact later in Section~\ref{sec:eval1}. A data sample is labeled by a target device if it contains at least one packet of the device (except for HomeMole that labels every packet in a window, as its paper describes), or labeled as negative if it only contains background traffic. Due to traffic fusion by middleboxes, a data sample is likely to contain more than one target device. The entire data samples are randomly separated by a ratio of 4:3:3 for training, validation and testing.

\textbf{Metrics.} Due to the imbalance of the mixed dataset, to express the accuracy of identification, we use 
1) \textit{precision}, i.e., the number of true positives divided by the number of predicted positives;
and 2) \textit{recall}, i.e., the number of true positives divided by the number of real positives. 
Given the vastly larger number of non-target devices in ISP networks, we also measure the false positive rate (FPR) to explore the impact of the base-rate fallacy~\cite{base-rate}. For runtime performance, we measure the processing time, throughput and runtime resource consumption of our framework.

\subsection{Identification Accuracy}
\label{sec:eval1}

\begin{figure}[t]
\centering
\subfigure[Toggle Devices]{
\begin{minipage}{0.46\linewidth}
\centering
\includegraphics[width=\textwidth]{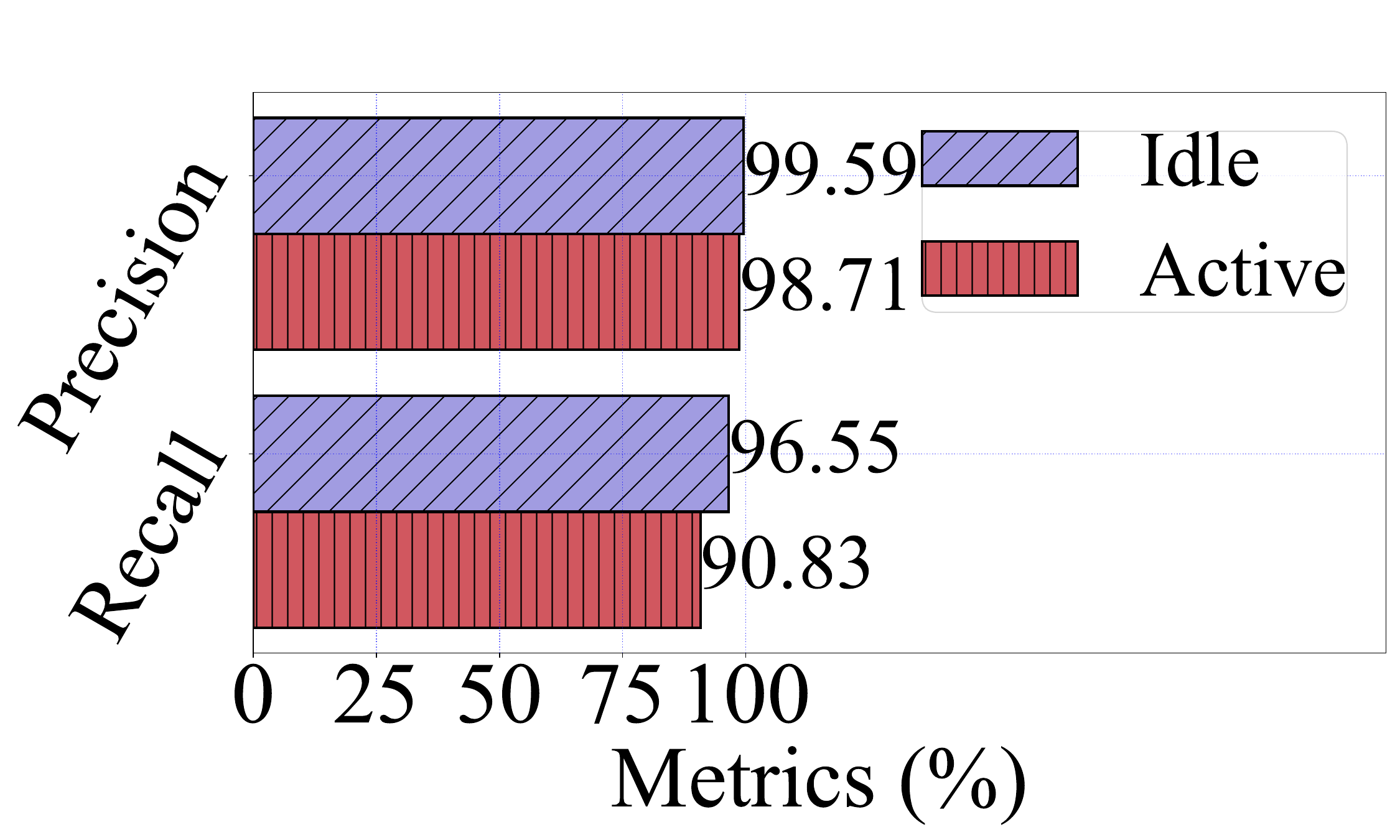}
\end{minipage}
}
\subfigure[Stream Devices]{
\begin{minipage}{0.46\linewidth}
\centering
\includegraphics[width=\textwidth]{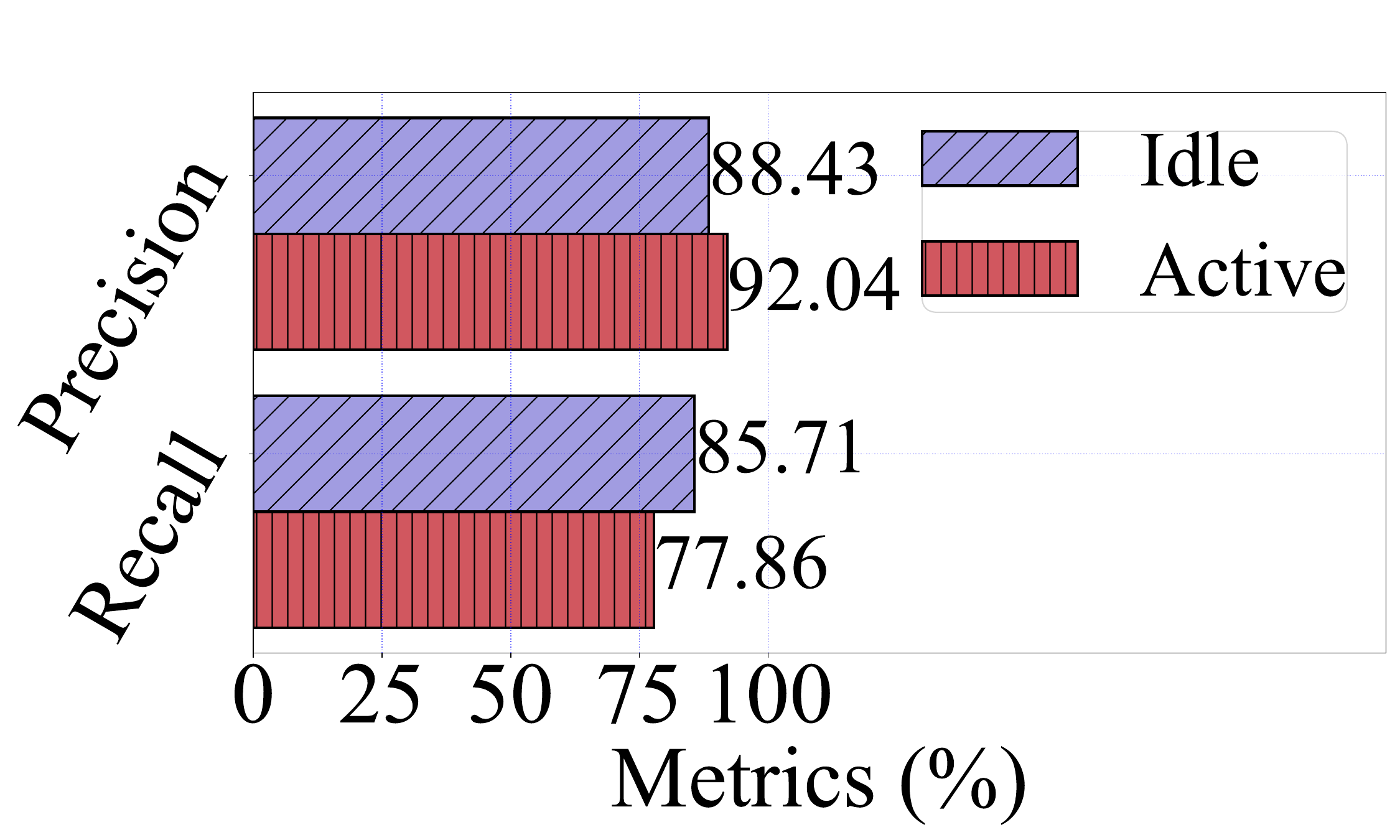}
\end{minipage}
}
\caption{Idle state versus active state.}
\label{fig:active}
\end{figure}

The result of device identification accuracy is shown in Fig.~\ref{fig:accuracy}, which illustrates the average metric value and the standard error in the four datasets.

In the NAT scenario, we find that DeviceRadar achieves remarkable average precision and recall over 90\% in all the datasets. 
Among the baselines, approaches like DeNAT and HomeMole also obtain good accuracy as they are specifically designed for the NAT scenario. Nonetheless, they use many more features than DeviceRadar, which can be unavailable in other middlebox scenarios like VPN. DarkSide shows the lowest accuracy, implying that the traffic statistics can be diluted by the high-speed background traffic in ISP networks and become unreliable for device identification. As for PingPong, though effective in identifying some of the devices, it shows a large fluctuation of accuracy across devices, mainly because it can only identify TCP-based devices but cannot identify purely UDP-based devices (e.g., an outlet in our testbed). In contrast, DeviceRadar exhibits the best stability of identification by the lowest standard errors in all the datasets. 

In the VPN scenario, DeviceRadar achieves the highest precisions and the highest recalls in all the datasets with even greater advantage over baseline methods. Compared to the effectiveness in the NAT scenario, DeNAT and HomeMole suffer from a degradation of over 60\% in the metrics.
This is due to the unavailability of many useful features in the VPN scenario (e.g., addresses used by DeNAT, port numbers used by HomeMole). In contrast, approaches like DeviceRadar and PingPong show almost no decline in accuracy as they only use the packet size and direction as features. In summary, we show that DeviceRadar can achieve higher and more stable accuracy of device identification than other methods when dealing with modified traffic by middleboxes like NAT and VPN.

We evaluate the influence of different base rates, i.e., the ratio of IoT traffic to background traffic, which can vary among links in ISP networks, on the metrics of precision, recall and FPR. As shown in Fig.~\ref{fig:base-rate}, all three metrics exhibit no significant changes with the increase of the base rate. Specifically, the FPR of DeviceRadar remains at a low level of 0.1\%, meaning that DeviceRadar has a low probability of misidentifying non-IoT traffic as IoT traffic of target vulnerable devices. It shows the good generality of DeviceRadar that can perform well at different links and networks of ISPs.

We are also curious about the effectiveness when the target devices are in the active state where specific functions are manually triggered via IoT apps. Based on the way of being activated,
we categorize IoT devices into \textit{toggle} devices (e.g., plugs-on/off) and \textit{stream} devices (e.g., cameras-watch). We write a script to continuously trigger certain functions of the devices in our testbed and collect the traffic mixed with the background traffic (details are described in the Appendix). 
In Fig.~\ref{fig:active}, we find that DeviceRadar retains good accuracy in both the device states, though the recall slightly drops in the active state, especially for stream devices. It is because stream devices typically generate a large number of packets that may not be key packets but are similar to the packets of large sizes in the background traffic (e.g., over 1000 bytes). Nonetheless, continuously using IoT apps is infrequent in real use, as most of the current IoT devices do not heavily rely on manual interactions but stay in a stable state most of the time. 

\begin{figure}[t]
\centering
\subfigure[Self-built]{
\begin{minipage}{0.46\linewidth}
\centering
\includegraphics[width=\textwidth]{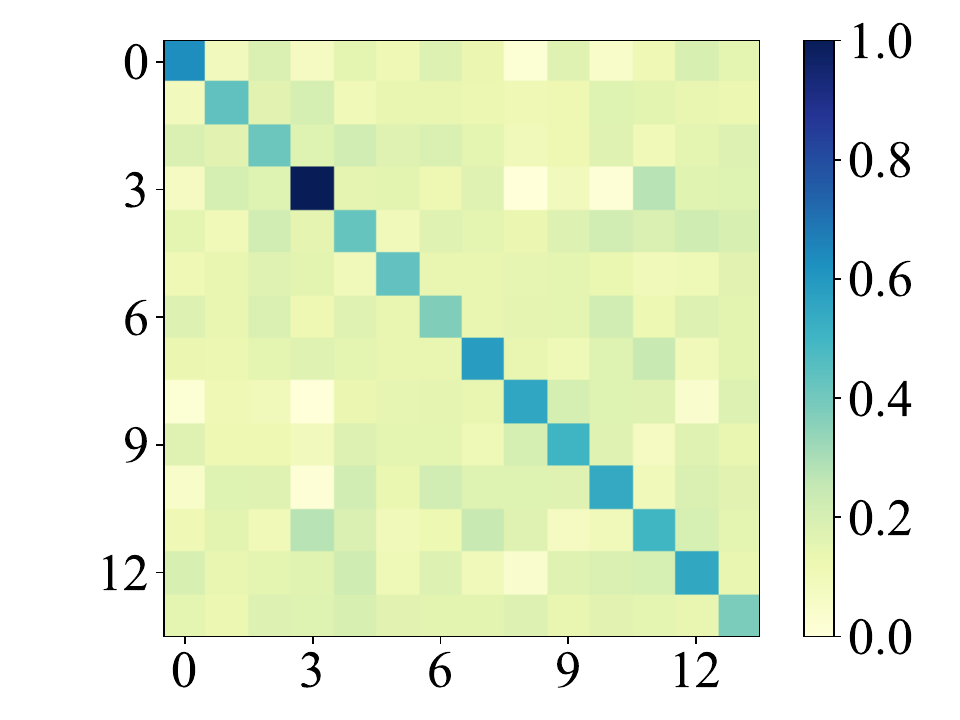}
\end{minipage}
}
\subfigure[NEU]{
\begin{minipage}{0.46\linewidth}
\centering
\includegraphics[width=\textwidth]{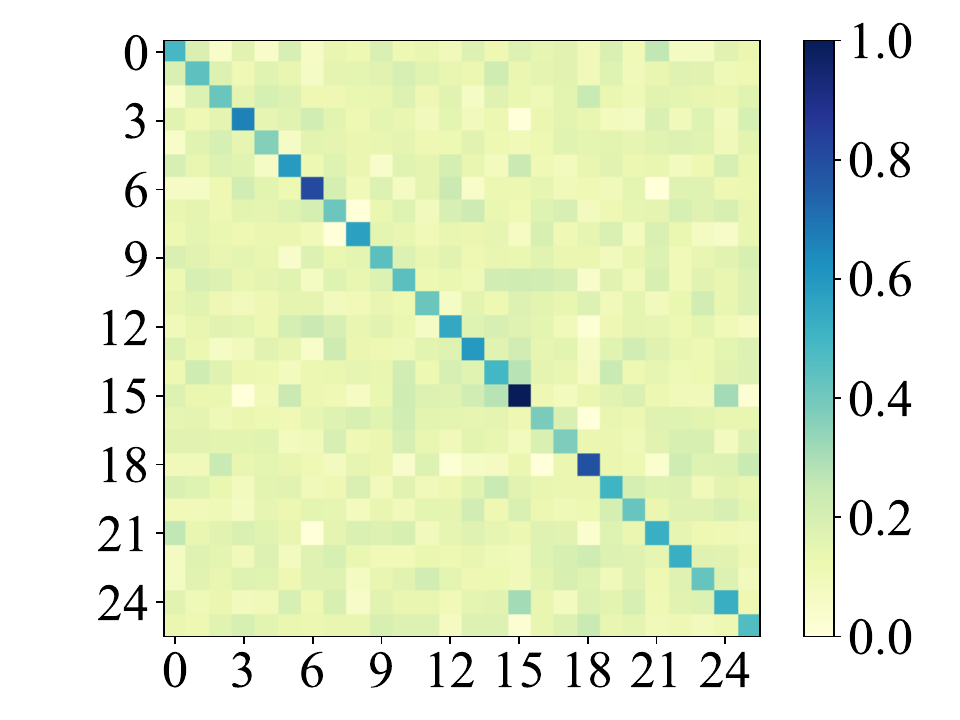}
\end{minipage}
}
\subfigure[ICL]{
\begin{minipage}{0.46\linewidth}
\centering
\includegraphics[width=\textwidth]{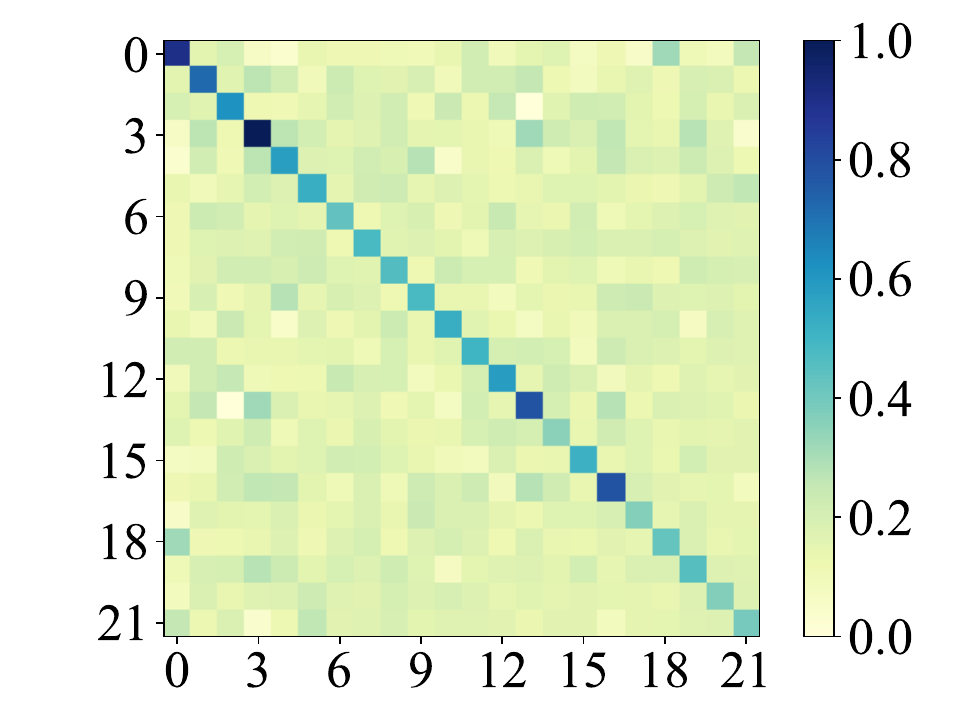}
\end{minipage}
}
\subfigure[UNSW]{
\begin{minipage}{0.46\linewidth}
\centering
\includegraphics[width=\textwidth]{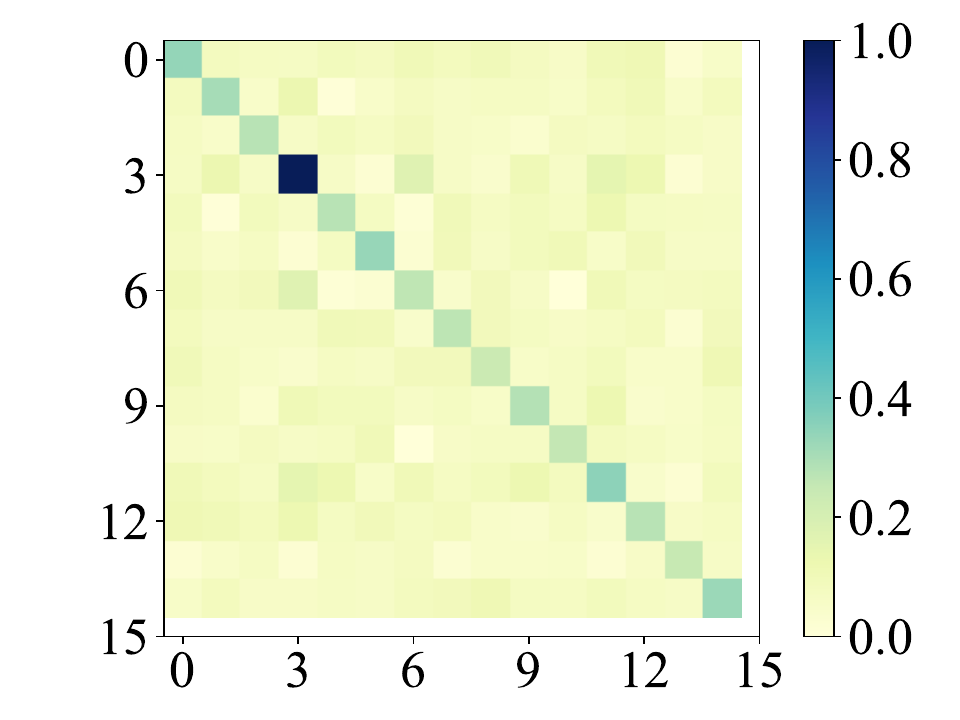}
\end{minipage}
}
\caption{Packet embedding similarity between devices.}
\label{fig:embed_sim}
\end{figure}

We believe the packet embedding is an important reason behind the high accuracy of DeviceRadar. To explore its effectiveness, we calculate the average cosine similarity between the packet embedding vectors from the same device and other devices. Fig.~\ref{fig:embed_sim} illustrates the matrix of packet embedding similarity between devices in each of the four datasets. We observe that the values on the diagonal, i.e., the packet embedding similarity within the same device is the largest for all the datasets and columns.
This means that our embeddings effectively predict the neighboring packets of the same device. In addition, we find that different devices of the same manufacturers show slightly higher similarities than devices of different manufacturers but are still distinguishable, such as device 3 (TPLink-camera) and device 6 (TPLink-plug) in UNSW, device 3 (MiAI-soundbox) and device 11 (Xiaomi-plug) in our testbed. It is because they share some of the general services among the devices of a brand, such as device access to the cloud (e.g., \url{devs.tplinkcloud.com:443}). Despite this, different types of devices have many other bursts of traffic for their specific functions (e.g., streaming of IP cameras), making our packet embedding still effective for distinguishing devices of the same brand.

\begin{figure}[t]
\centering
\subfigure[Precision]{
\begin{minipage}{0.46\linewidth}
\centering
\includegraphics[width=\textwidth]{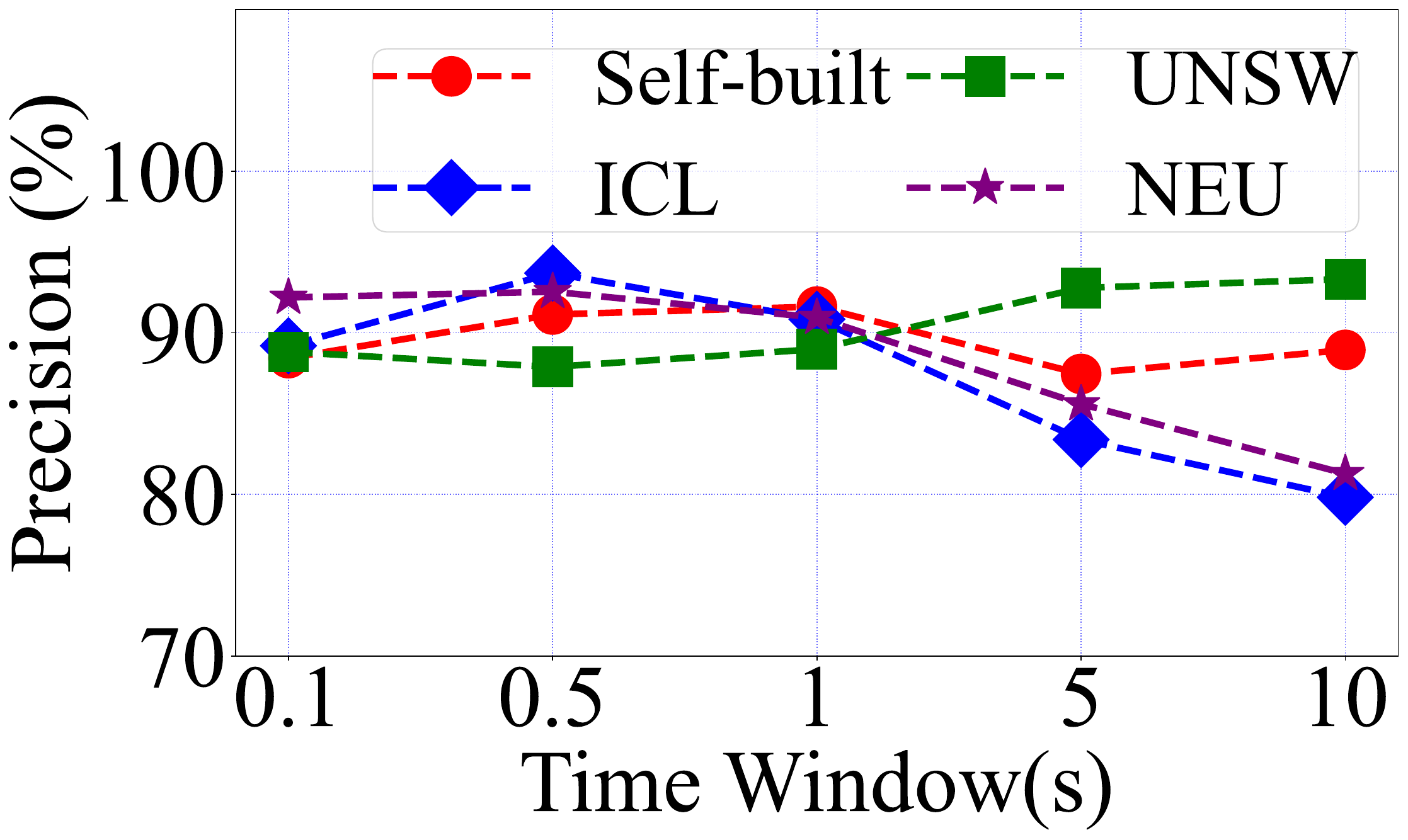}
\end{minipage}
}
\subfigure[Recall]{
\begin{minipage}{0.46\linewidth}
\centering
\includegraphics[width=\textwidth]{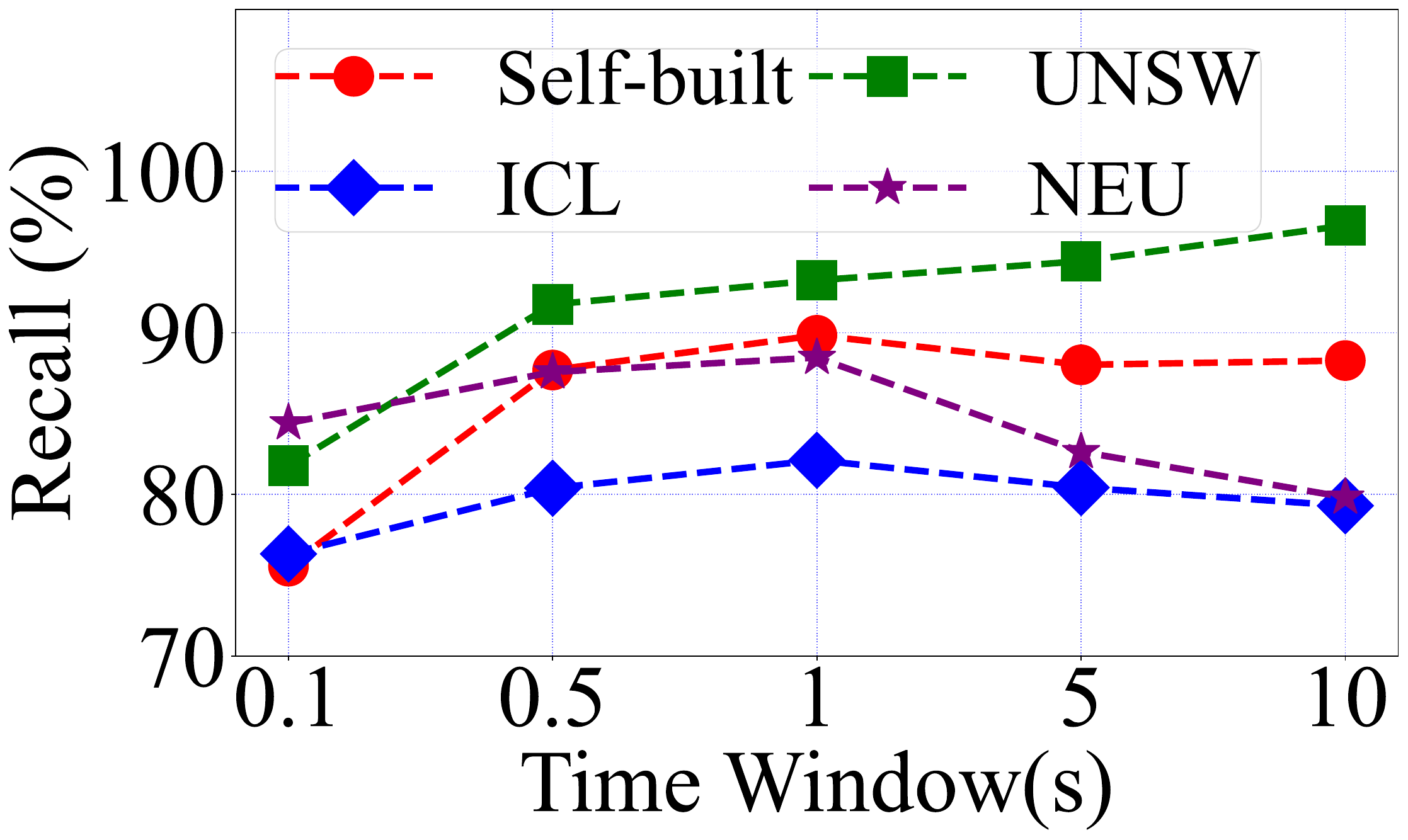}
\end{minipage}
}
\caption{Sensitivity experiment on time window size.}
\label{fig:hyper}
\end{figure}

\begin{table}[t]
    \centering
    \caption{\textcolor{black}{Identification accuracy across various categories.}}
    \begin{tabular}{c|ccc|ccc}
    \toprule
    \multirow{2}{*}{\textcolor{black}{Category}}  & 
\multicolumn{3}{c|}{\textcolor{black}{PingPong}} & \multicolumn{3}{c}{\textcolor{black}{DeviceRadar}} \\
    \cline{2-7}
    ~ & \textcolor{black}{Rec} & \textcolor{black}{Prec} & \textcolor{black}{FPR} & \textcolor{black}{Rec} & \textcolor{black}{Prec} & \textcolor{black}{FPR} \\
    \hline
    \textcolor{black}{Automation} & \textcolor{black}{0.488} & \textcolor{black}{0.541} & \textcolor{black}{0.266} & \textcolor{black}{\textbf{0.851}} & \textcolor{black}{\textbf{0.925}} & \textcolor{black}{\textbf{0.0070}} \\
    \textcolor{black}{Camera} & \textcolor{black}{0.629} & \textcolor{black}{0.231} & \textcolor{black}{0.726} & \textcolor{black}{\textbf{0.883}} & \textcolor{black}{\textbf{0.880}} & \textcolor{black}{\textbf{0.011}} \\
    \textcolor{black}{TV/media} & \textcolor{black}{\textbf{0.960}} & \textcolor{black}{0.152} & \textcolor{black}{0.767} & \textcolor{black}{0.586} & \textcolor{black}{\textbf{0.757}} & \textcolor{black}{\textbf{0.0038}} \\
    \textcolor{black}{Speaker} & \textcolor{black}{\textbf{0.987}} & \textcolor{black}{0.242} & \textcolor{black}{0.705} & \textcolor{black}{0.866} & \textcolor{black}{\textbf{0.872}} & \textcolor{black}{\textbf{0.0055}} \\
    \textcolor{black}{Hub} & \textcolor{black}{0.843} & \textcolor{black}{0.335} & \textcolor{black}{0.595} & \textcolor{black}{\textbf{0.888}} & \textcolor{black}{\textbf{0.945}} & \textcolor{black}{\textbf{0.0013}} \\
    \textcolor{black}{Appliance} & \textcolor{black}{0.0021} & \textcolor{black}{0.604} & \textcolor{black}{0.0002} & \textcolor{black}{\textbf{0.750}} & \textcolor{black}{\textbf{0.800}} & \textcolor{black}{\textbf{0.0001}} \\
    \textcolor{black}{Router} & \textcolor{black}{0.919} & \textcolor{black}{0.509} & \textcolor{black}{0.378} & \textcolor{black}{\textbf{0.952}} & \textcolor{black}{\textbf{0.918}} & \textcolor{black}{\textbf{0.0061}} \\
    \bottomrule
    \end{tabular}
    \label{tab:categories}
\end{table}

We also conduct a sensitivity experiment on an important hyperparameter of DeviceRadar: the time window size. Typically, an overly short window cannot capture sufficient information, while an overly long window is not suitable for real-time use and also contains too many background packets that may dilute the traffic of the target devices. The results are shown in Fig.~\ref{fig:hyper}. The precision decreases after 1 second except for on the dataset of UNSW; the recall grows or remains nearly unchanged with the increase of the time window, except for on the dataset of NEU in which the recall drops after 1 second. Overall, DeviceRadar can obtain a relatively better accuracy when the time window is set to 1 second. The result also suggests that DeviceRadar can conduct an inference on the existence of target devices for every second, which lays a foundation for realizing real-time device identification.

\textcolor{black}{
Finally, to assess identification differences across device categories, we employ a more comprehensive benchmark~\cite{sok}, featuring 45 smart home devices. 
It includes 10 cameras, 7 TVs/media devices, 7 speakers, 6 hubs, 2 home routers, 2 appliances, and 11 home automation devices such as plugs, smoke alarms, and garage openers. We utilize the same preprocessing steps as in prior experiments and include the baseline PingPong for comparison.
Table~\ref{tab:categories} presents the recall, precision, and FPR across these categories. DeviceRadar exhibits improved recall for most categories and better precision and FPR across all categories, except for TVs and speakers. The results of the baseline with these categories are due to its treatment of their frequent streaming packets of MTU size as signatures, which are also common in background traffic (see Fig.~\ref{fig:mawi}), leading to numerous false positives. DeviceRadar, relying on the proposed key packets, maintains much higher precision and lower FPR, highlighting its better usability.
}

\subsection{Runtime Performance}
\label{sec:eval2}

\begin{figure}[t]
\centering
\subfigure[Processing latency]{
\begin{minipage}{0.46\linewidth}
\centering
\includegraphics[width=\textwidth]{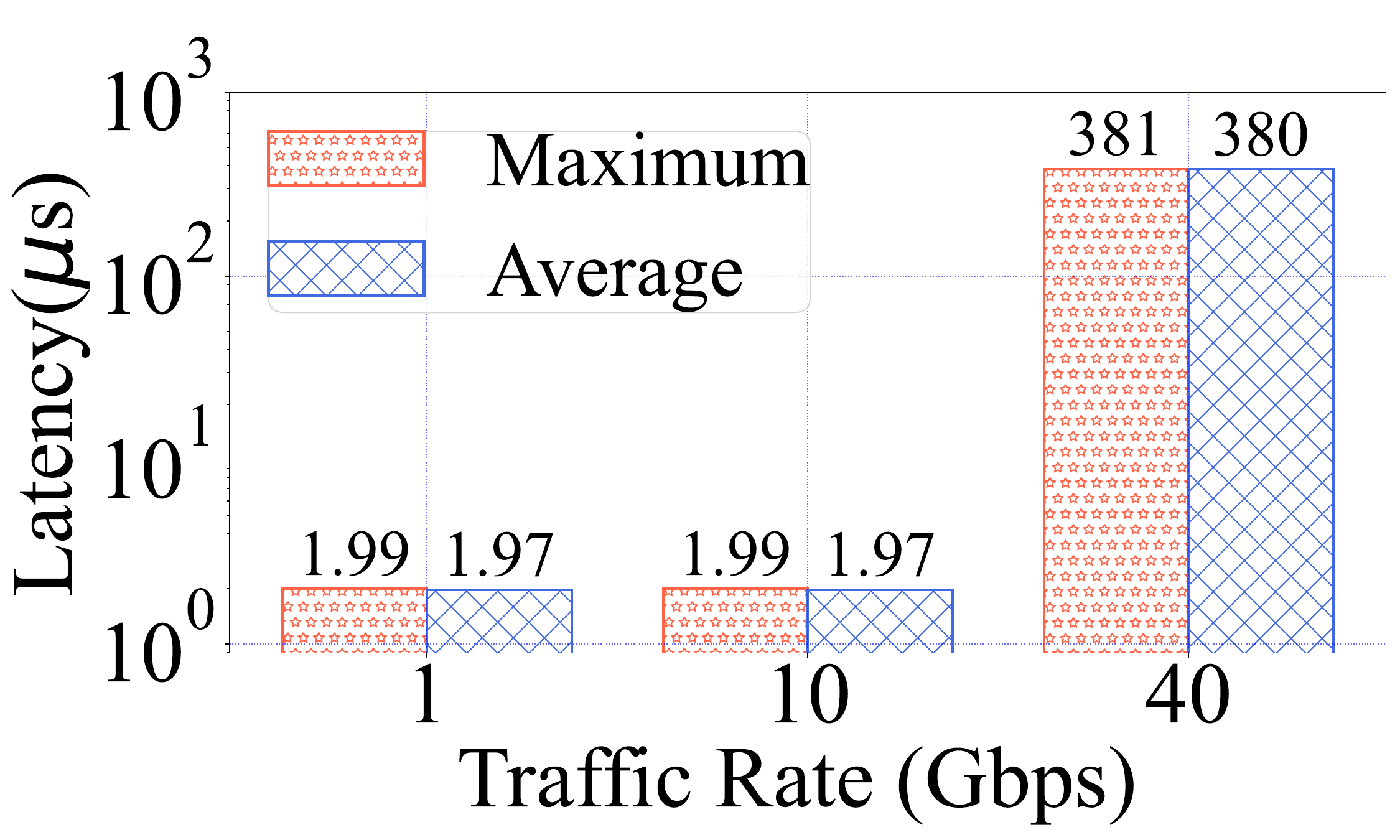}
\end{minipage}
}
\subfigure[Throughput]{
\begin{minipage}{0.46\linewidth}
\centering
\includegraphics[width=\textwidth]{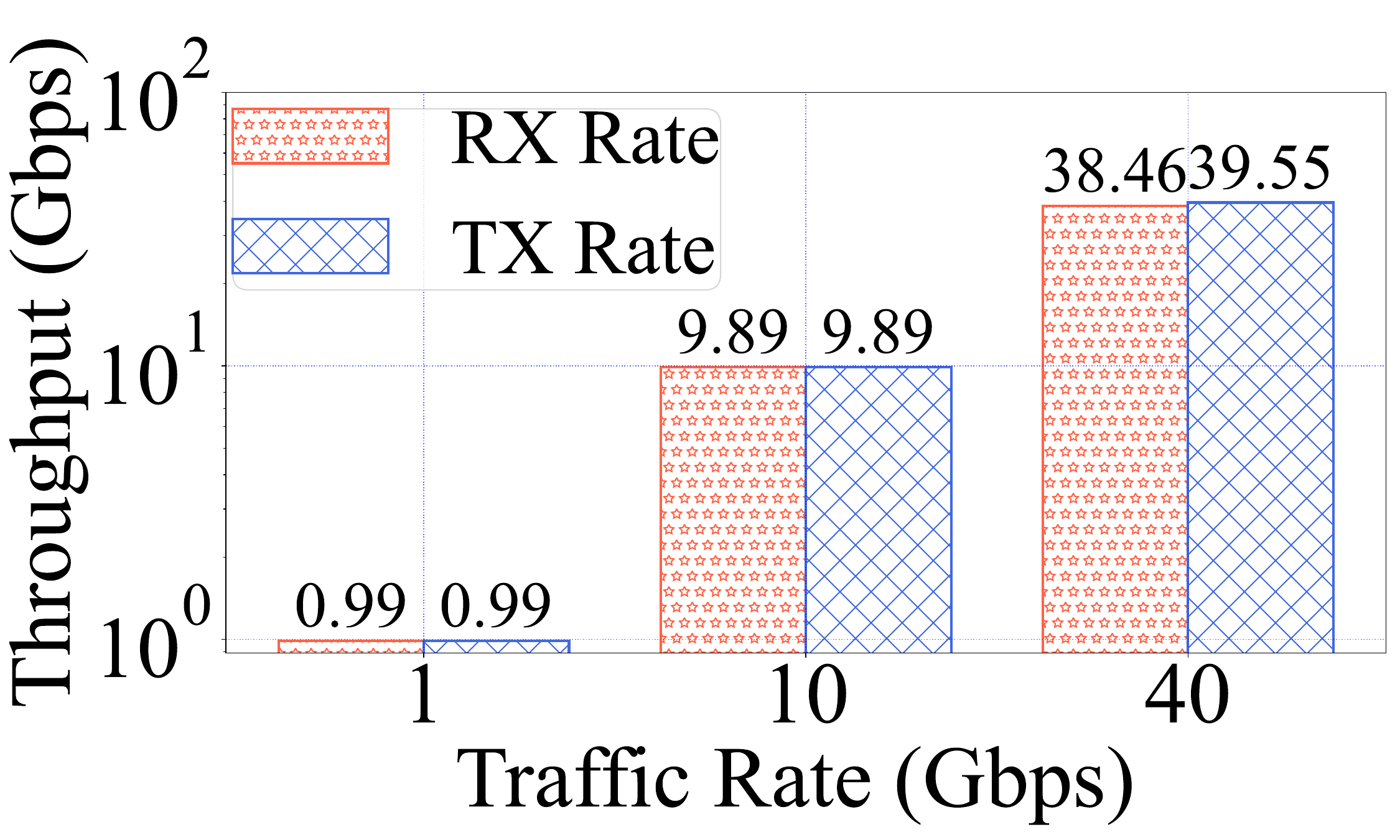}
\end{minipage}
}
\caption{Processing latency and throughput of DeviceRadar.}
\label{fig:runtime}
\end{figure}

To test the runtime performance of DeviceRadar, we use the network tester to send our mixed traffic at rates of 1 Gbps, 10 Gbps and 40 Gbps. 
These rates simulate the real-world traffic speeds on an ISP edge switch, an ISP aggregation/core switch and a high-speed ISP core switch under the stress of line rate, respectively. We illustrate the results in Fig.~\ref{fig:runtime}. We observe that, when the traffic rate is set to 1 Gbps and 10 Gbps, DeviceRadar is able to process packets at a very high speed of under 2 microseconds (1 microsecond = $1\times 10^{-6}$ second). 
When the traffic rate is set to the line speed (40 Gbps), the processing latency increases to 380 microseconds on average (which is still suitable for line speed). Compared to the time window for one inference (1 second), this processing latency is trivial and is qualified for real-time device fingerprinting. 
As for throughput, thanks to the high performance of the programmable switch, DeviceRadar can achieve nearly no packet loss and realize high throughput no matter how the traffic rate varies. These results show that, by managing to deploy the device fingerprinting components on the programmable data plane, DeviceRadar achieves real-time and high-throughput device identification in ISP networks. 

To measure the runtime overhead of the baselines, we adopt the SDN deployment shown in Fig.~\ref{fig:deploy}, where the SDN switch is implemented by a server running Open vSwitch (OvS) and the controller runs the baseline approaches, and they are placed within the same local network. As these approaches are not specific for online use, to simulate the online testing, we use the OvS server to report the extracted feature vectors to the controller one by one; on the controller, the testing batch size of the baseline approaches is set to 1. The ML-based method DarkSide is measured on a CPU, and the DL-based method HomeMole is measured on both a CPU and a GPU. 

Table~\ref{tab:time_compare} shows the window size for one inference (i.e., one time of device identification), communication latency and inference time. First, because of the fully in-network implementation, DeviceRadar completely eliminates the communication latency, which highlights its advantage for online deployment. Further, as the inference process of DeviceRadar is directly embedded in the pipeline of packet processing (i.e., by the inference table introduced in Section~\ref{subsec:mat}), its inference time is equal to the packet processing latency, which is in microseconds. It is far less than the inference time of the baselines, which is in milliseconds even with the acceleration of GPUs. As for the window size, DeviceRadar can make an inference to identify target devices for every one second. Compared to some existing approaches discussed in Section~\ref{sec:limits} that require hours of analysis for one inference (e.g., DNS-based approaches), the identification result of DeviceRadar is much more timely and thus more valuable to subsequent prompt responses. Other baselines may use a spatial window of packets. For example, the window size of DarkSide is 180 packets. Given the rate in the background traffic of 600 Kpps on average, it needs to have the ability to make an inference every 0.3 milliseconds. As the window is even shorter than the communication latency or inference time, it suggests that the inference process will never catch up with the online traffic. In summary, DeviceRadar significantly outperforms other methods in terms of online overhead, showing its great practicality for online device fingerprinting in ISPs.


\begin{table}[t]
    \footnotesize
    \centering
    \caption{Comparison of processing time.}
    \label{tab:time_compare}
    \begin{tabular}{ccccc}
    \toprule
    \makecell[c]{Attribute} & \makecell[c]{DarkSide \\ (ML-based)} & \multicolumn{2}{c}{\makecell[c]{HomeMole \\ (DL-based)}} & \makecell[c]{DeviceRadar} \\
    \midrule
    \makecell[c]{Running \\ device} & CPU & CPU & GPU & \makecell[c]{Switch \\ ASIC} \\
    \midrule
    \makecell[c]{Window \\ size} & \makecell[c]{0.3 ms \\ (180 pkts)} & \makecell[c]{0.167 ms \\ (100 pkts)} & \makecell[c]{0.167 ms \\ (100 pkts)} & 1 sec \\
    \midrule
    \makecell[c]{Comm. \\ latency (ms)} & 3.81 & 6.53 & 6.53 & 0.0 \\
    \midrule
    \makecell[c]{Inference \\ time (ms)} & 10.2 & 45.3 & 0.152 & \makecell[c]{1.97 $\times 10^{-3}$} \\
    \midrule
    \makecell[c]{Online \\ practical?} & No & No & No & Yes \\
    \bottomrule
    \end{tabular}
\end{table}

\textcolor{black}{
The resource consumption on the data plane (or runtime resource consumption) is measured by the computing resources, including the exact match crossbar (eMatch xBar, used by match processes), VLIW (Very Long Instruction Word, used by action processes), and memory resources including logical table ID, map RAM and SRAM.
We compare the resource consumption of basic switching functions with the combination of DeviceRadar and these functions. The P4 program for basic switching functions derived from Tofino supports L2 switching, IPv4 switching, IPv6 switching, VLAN port mapping, and Segment Routing over IPv6 (SRv6). The results, depicted in Fig.~\ref{fig:usage}, reveal that the resource overhead introduced by DeviceRadar is relatively modest. The overall resource usage remains within acceptable limits, supporting both switching functions and DeviceRadar. This demonstrates the feasibility of deploying DeviceRadar on the data plane.
}


\subsection{Use Case}
\label{sec:eval3}

Given the high accuracy, real-time detection and high throughput, DeviceRadar provides timely knowledge about the existence of certain IoT devices in the network, which can be conveniently integrated with other network tasks. We highlight this by presenting a use case -- the integration with the mitigation of DDoS attacks in IoT. There are existing works on the efficient detection of malicious traffic including IoT botnets and DDoS attacks, of which the typical defense logic is to take actions (e.g., throttling, blocking) \textit{after} detecting the ongoing attack traffic. In contrast, DeviceRadar makes a new solution possible, which is to timely identify the vulnerable devices \textit{before} they are compromised and being utilized to launch DDoS attacks. These vulnerable devices can be found on well-established resources like CVE~\cite{cve} which describe their exploits and behaviors, so that we can actively prevent them immediately when they are identified.

\begin{figure}[t]
\centering
\begin{minipage}{0.48\linewidth}
    \centering
    \includegraphics[width=\linewidth]{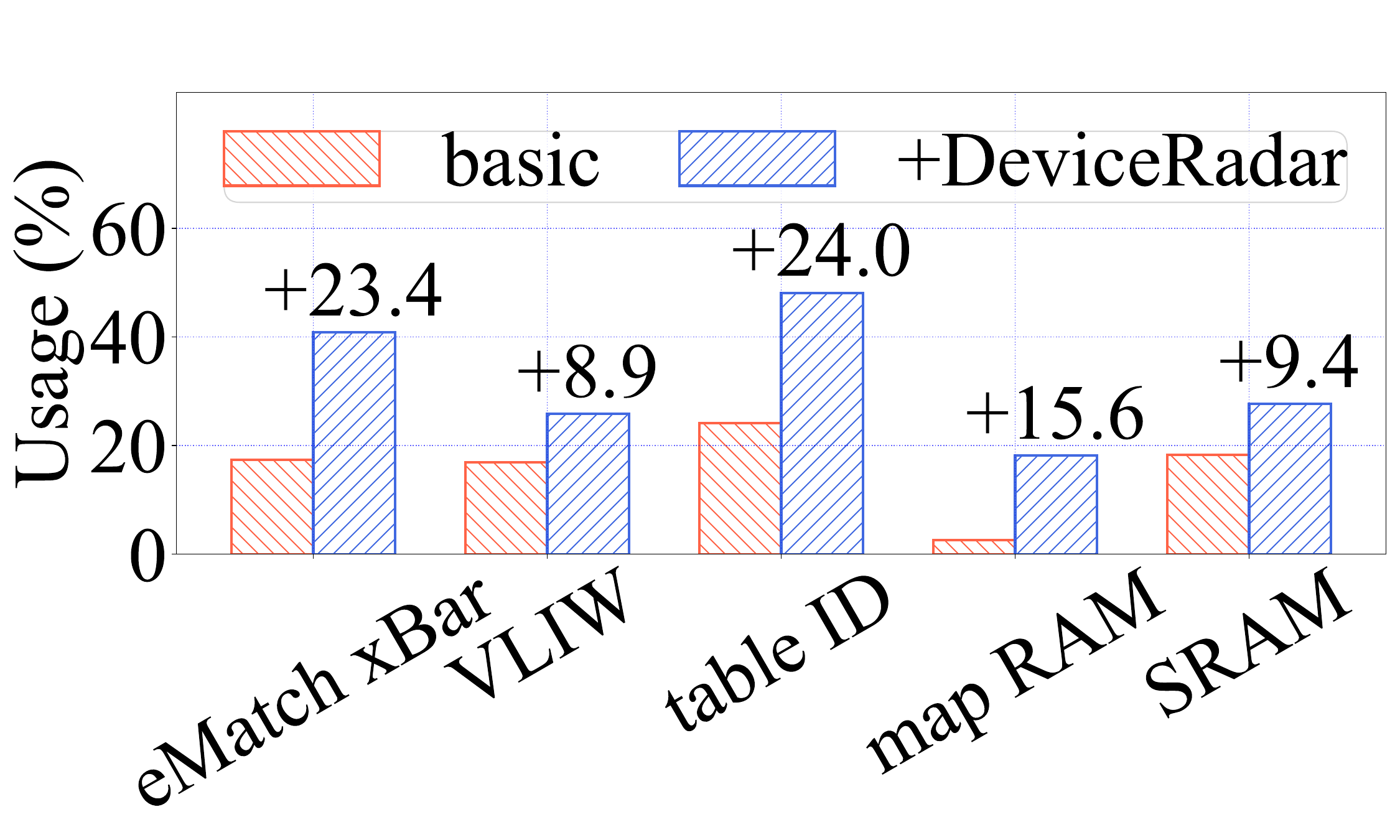}
    \caption{\textcolor{black}{Resource usage versus basic switching function.}}
    \label{fig:usage}
\end{minipage}
\begin{minipage}{0.48\linewidth}
    \centering
    \includegraphics[width=\linewidth]{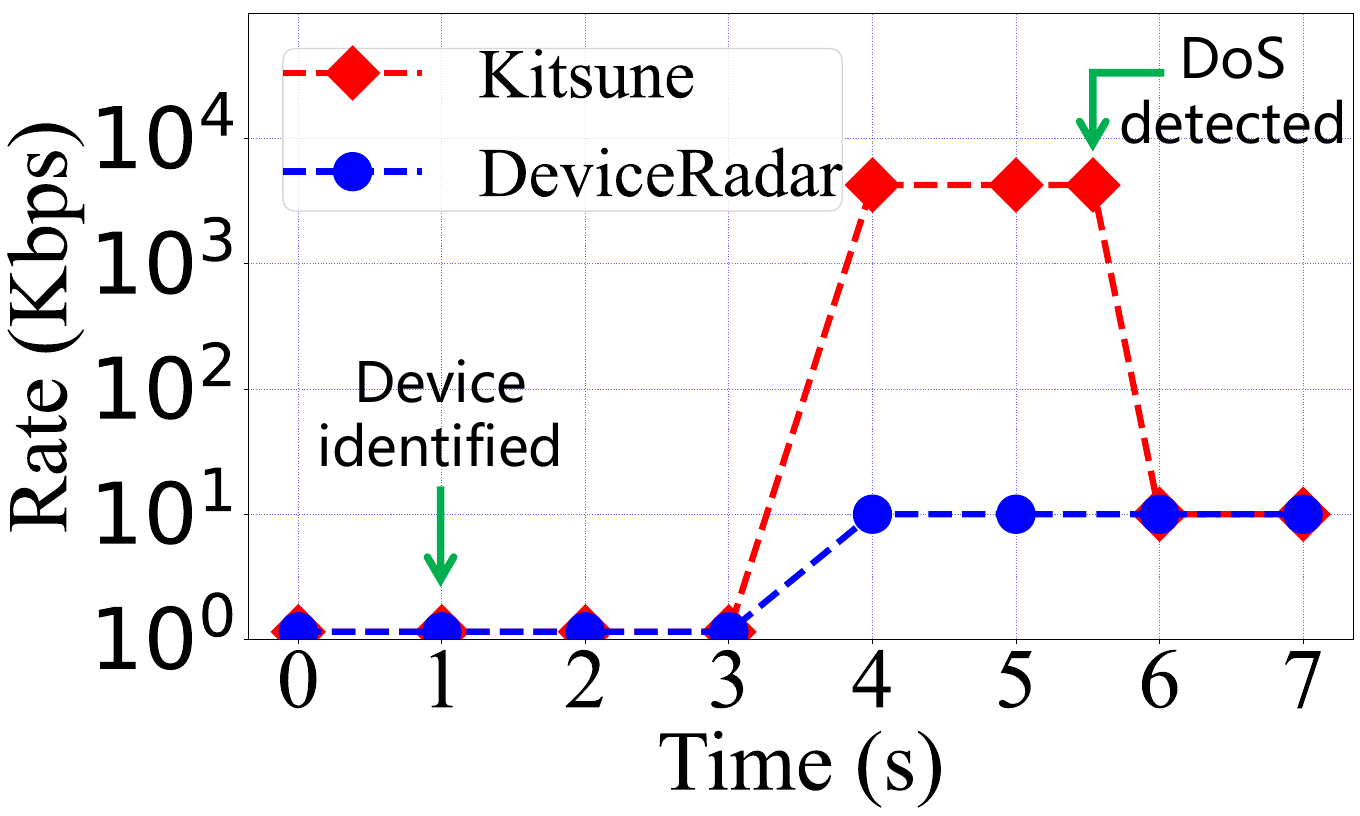}
    \caption{Use case: integration with DDoS mitigation.}
    \label{fig:ddos}
\end{minipage}
\end{figure}

To demonstrate this use case, we use a Raspberry Pi to behave as a bulb that periodically synchronizes its property with a cloud, which can be implemented by the virtual device of AWS IoT Greengrass~\cite{grengrass}. Its normal behavior is to send a packet of 107 bytes out and receive a packet of 40 bytes per second (rate: 1.15 Kbps on average). We assume it is known to be vulnerable to IoT malware like Mirai that exploits bots to launch DDoS attacks, which should be paid attention to. First, we let DeviceRadar treat this ``bulb'' as a target device. Once the device is identified, a rule is deployed for protection in advance: limiting the rate of this device to a very low level of 10 Kbps. Then we simulate a quick attack scenario, in which after three seconds of connecting to the network, the ``bulb'' gets compromised and starts to send flooding traffic at 20,000 packets per second. For comparison, we employ Kitsune~\cite{kitsune}, a state-of-the-art lightweight intrusion detection system. Once the attack traffic is detected, Kitsune uses the same rule of traffic throttling for mitigation.

As shown in Fig.~\ref{fig:ddos}, DeviceRadar can identify the device in real-time (using about 1 second) so that the mitigation policy is timely pre-set before the attack. In contrast, Kitsune has to wait for the attack reaching to a significant level that can be detected (using about 1 second) and then process the inference (using 0.54 seconds). 
\textcolor{black}{
As such, there are almost 2 seconds in which the attack reaches its peak of 4.27 Mbps. Do not underestimate its impact: if we consider a large-scale botnet like Mirai that controls over 145,000 devices to flood one victim~\cite{mirai}, such a DDoS attack can reach a peak of over 604.63 Gbps for seconds, sufficiently resulting in disastrous consequences. 
}
It shows that a real-time device fingerprinting method like DeviceRadar is greatly helpful to the promotion of the integrated defense system.

\textcolor{black}{
We also test the feasibility of deploying such massive per-device defense rules for identified vulnerable devices in terms of resource costs. Among the 48 stages of a Tofino switch, we leave only one stage for applying per-IP rules. Our result shows that the SRAM resources of a stage can support the installation of up to 360,000 per-IP rules. In comparison, the number of source IP addresses in our WIDE backbone trace is 86,255, suggesting that the resource of one stage is adequate to maintain over 4 rules even for each host at a vantage point in an ISP network. Therefore, it is practical to deploy sufficient defense rules with a trivial memory overhead.
}

\begin{figure}[t]
\centering
\begin{minipage}{0.46\linewidth}
    \centering
    \includegraphics[width=\linewidth]{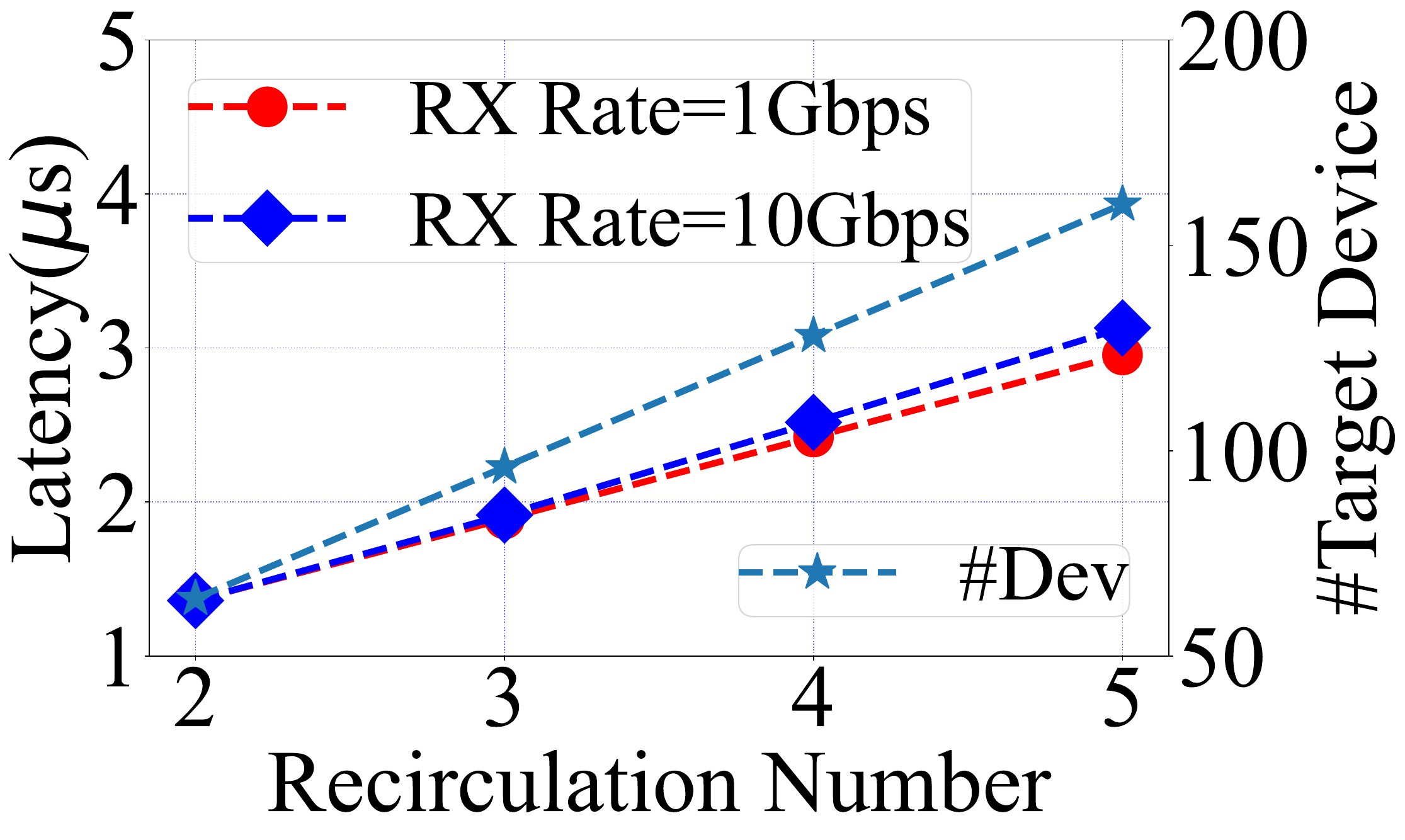}
    \caption{Scalability.}
    \label{fig:scalability}
\end{minipage}
\begin{minipage}{0.46\linewidth}
    \centering
    \includegraphics[width=\linewidth]{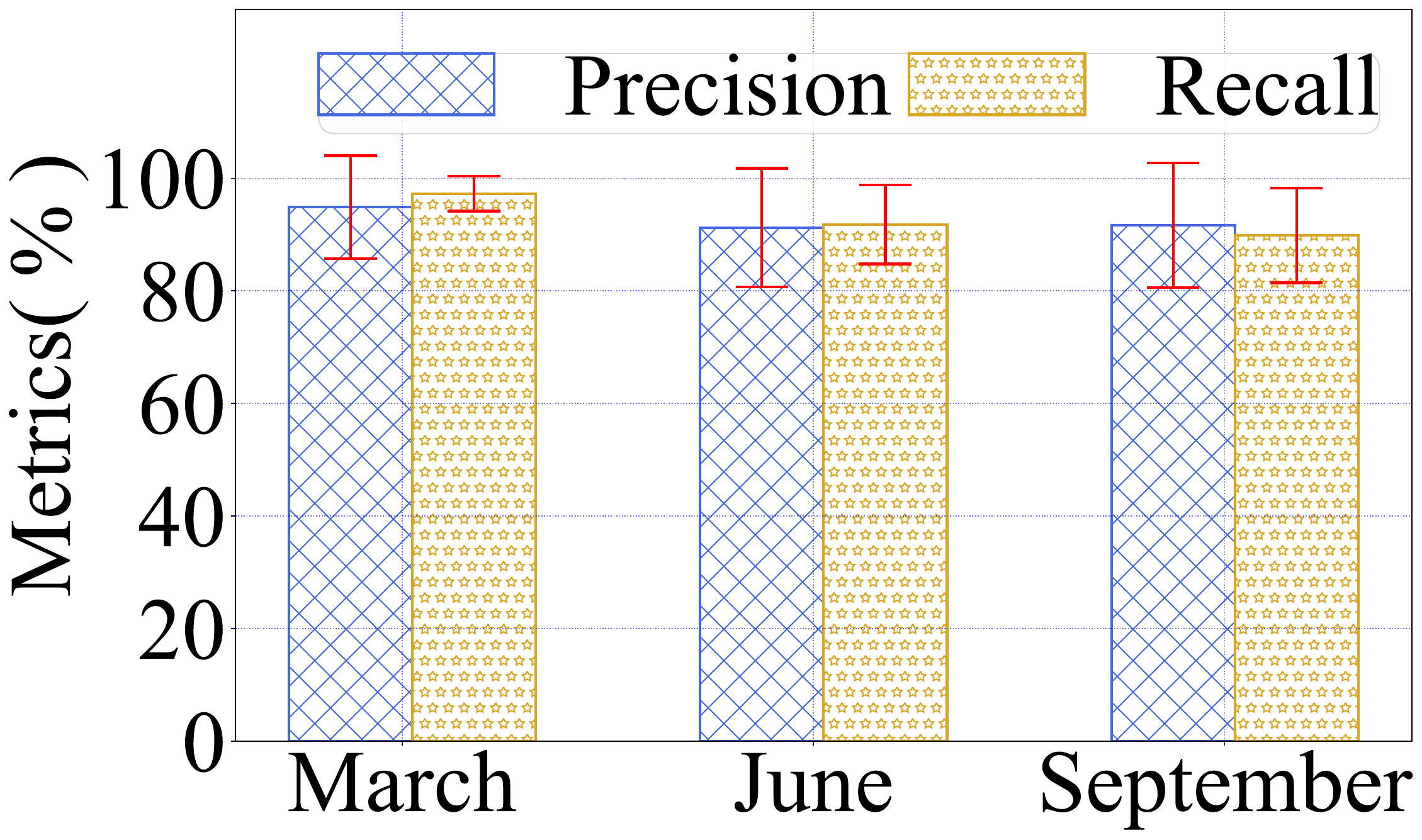}
    \caption{Stability.}
    \label{fig:stability}
\end{minipage}
\end{figure}

\subsection{Discussion}
\label{sec:discussion}

\textbf{Scalability.} Another constraint of P4 switches is the limited number of match-action stages. For example, a Tofino switch supports a maximum of 48 stages by concatenating 4 pipelines of 12 stages each. Since DeviceRadar generates match-action tables per target device type, the number of device types to be identified in our implementation is limited to 32 on a switch. For better scalability, we can use a technique called \textit{recirculation} that re-sends the egress packet to the ingress port to utilize more match-action stages. As Fig.~\ref{fig:scalability} shows, it can linearly increase the supported number of device types up to 160. Though the processing latency also increases, it is still at a level of microseconds, which is sufficient for online use.
Besides, an ISP typically only needs to focus on a small set of device types that 1) are found vulnerable to prevalent attacks and 2) have a large use percentage. For example, Torabi et al.~\cite{investigate} find that no more than 4 IoT device types account for 99.4\% of all of the compromised consumer IoT devices.

\textbf{Stability to IoT updates.} Due to possible updates of firmware, software and servers of IoT, the efficacy of device fingerprinting may suffer from the issue of concept drift and becomes unstable over time. Thus, we use the datasets of March, June and September to assess the impact. During this period, we observed that most of the devices have received at least one update. The datasets are mixed with the background traffic to evaluate the original model constructed in March. Fig.~\ref{fig:stability} illustrates the result. We see that the decline in accuracy is trivial even after six months. In view of such a long period, DeviceRadar has good stability over time.

\textcolor{black}{
\textbf{Newly added devices.}
False alarms may occur when unseen devices, such as brand-new products, join a network and are misidentified by models. 
To test this, we perform an experiment assessing this scenario.
We remove one device from the training dataset to simulate an unseen device, and train models for the remaining devices as targets.
We then test the model on a test set including the unseen device, and evaluate the false alarms on the removed device for each model. The average false alarm rate is 0.0201 $\pm$ 0.0071. 
While this rate is low, it indicates the potential for false positives in such situations, suggesting a limitation of our approach. To address this concern, it is necessary to perform periodic retraining by incorporating new devices. 
}

\textbf{Packet padding.} As DeviceRadar handles middlebox scenarios by only using packet sizes and directions as features, it is naturally susceptible to packet padding techniques such as padding to MTU~\cite{padding}.
However, we do not observe any IoT devices in any of the datasets using such padding techniques, possibly because the introduced overhead departs from the general design principle of IoT (that prioritizes lightweight and stable connections). Besides, packet padding can also be implemented on home gateways by users, typically because the users do not trust the ISPs and are reluctant to release any of their device information~\cite{spy}. As a security framework for ISPs, DeviceRadar respects their willingness but meanwhile cannot provide any related services, such as notifying the users of high-risk devices in their residences. For IoT botnet adversaries, given that one of their main purposes is to launch high-speed DDoS attacks, the benefit of using the padding techniques will not outweight the loss of attack effectiveness due to considerable network delay.

\textcolor{black}{
\textbf{Potential adversaries.}
Our approach only uses limited available features in the middlebox scenarios. Thus, adversaries may command their IoT bots to mimic untargeted devices (e.g., modifying packet sizes) to avoid detection. 
However, we argue that such adversaries may encounter difficulties in practice.
As our framework can identify target devices in real-time and promptly deploy defense rules, adversaries will find it difficult to compromise an IoT device using a known vulnerability.
This makes it difficult to command the device to mimic other devices.
In other cases, a vulnerable device may have been compromised before DeviceRadar is installed.
Despite this, we emphasize that adversaries cannot control packets sent from the IoT cloud.
This feature is a key part of our device fingerprinting, enabling us to still detect compromised devices even if they are trying to hide their activity.
Besides, tampering with the packet sizes may harm the integrity of the packets verified by the cloud side and affect normal device functions. In this case, IoT users may reboot or disconnect the devices, causing the adversaries to lose control.
}



\section{Conclusion and Future Work}
This paper proposes DeviceRadar, an online IoT device fingerprinting framework for ISP networks.
We realize accurate device identification even when the traffic is modified by middleboxes like NATs and VPNs, and deploy the proposed models on a programmable switch for online use with line-speed processing.
\textcolor{black}{
Our evaluation reveals that DeviceRadar achieves over 90\% identification accuracy with middleboxes for various IoT device types at a throughput of 40 Gbps, using microsecond-level processing latency. This accounts for only 1.3\% of GPU-based solutions.
}
For future work, we plan to evaluate DeviceRadar on more middleboxes, such Tor nodes and other VPN protocols like L2TP. 
\textcolor{black}{
From a more practical point of view, we wish to explore ways of porting DeviceRadar to a multi-switch environment. 
One possible way is to deploy DeviceRadar on each switch and detect IoT devices separately. This would need an additional algorithm that monitors the characteristics of the links and properly integrates the detection results from multiple switches. Another solution is to deploy one detection model in a distributed manner by splitting a decision tree into multiple sub-trees. In this way, selecting the best switches for deployment is an open issue, for which we may explore service chain optimization approaches to determine the organization of the switches.
}

\bibliographystyle{IEEEtran}
\bibliography{bibliography}


\end{document}


\title{Supplementary Material: Appendix}

\maketitle

\section{Proof of Claim}
\label{sec:proof}
The training objective of the packet embedding can also be interpreted as to maximize the probability of predicting the context for semantically similar packets. For $e_i,e_j$ that are embedding vectors of neighboring packets, $e_i$ and $e_j$ are trained to predict similar contexts denoted by $c$. We use $\epsilon$ to denote the similarity of two similar vectors and we have the following equation:
\begin{equation}
\begin{aligned}
    & e_i^T \cdot c - e_j^T \cdot c = \epsilon \\
    & \iff  (e_i - e_j) \cdot c^T = \epsilon \\ 
    & \iff (e_i - e_j) \cdot c^T \cdot c \cdot (e_i^T - e_j^T) = {||\epsilon||}^2 \\
    & \iff (e_i^T - e_j^T)\cdot (e_i - e_j) = \frac{{||\epsilon||}^2}{{{||c||}^2}} \\
    & \iff {||e_i||}^2 + {||e_j||}^2 - 2e_i^T\cdot e_j = \frac{{||\epsilon||}^2}{{{||c||}^2}}
\end{aligned}
\end{equation}

The cosine similarity between $e_i$ and $e_j$ can be calculated by the following equation:
\begin{equation}
\begin{aligned}
    & \mathrm{cosine}(e_i,e_j) = \frac{e_i^T \cdot e_j}{||e_i|| \cdot ||e_j||} = \frac{{||e_i||}^2 + {||e_j||}^2 - \frac{{||\epsilon||}^2}{{{||c||}^2}}}{2||e_i|| \cdot ||e_j||} 
\end{aligned}
\end{equation}

We set $\epsilon \leftarrow 0$, which means the context of $e_i$ and $e_j$ is sufficiently similar, and we have:
\begin{equation}
\mathrm{cosine}(e_i, e_j) \leftarrow \frac{{||e_i||}^2 + {||e_j||}^2}{2||e_i|| \cdot ||e_j||} \geq 1
\end{equation}

Since the value of the cosine similarity is in $[-1, 1]$, we have $\mathrm{cosine}(e_i,e_j) = 1$, which means their cosine similarity is theoretically maximized.

\section{Testbed Information}
\label{subsec:testbed}

We use two wireless routers as access points (AP) to avoid the excessive connections on a single AP that may cause disconnection. Note that these two APs do not enable any routing or DHCP functions; they just function like L2 switches. The two APs are connected to an enterprise router that assigns IP addresses to each connected device and also supports middlebox functions like NAT and VPN. We mirror the traffic of the two ports connected with the two APs to another port connected with a computer, which runs a tcpdump script to capture the traffic. External hard drives are used for the storage of collected traffic.

\begin{figure}[h]
\centering
\subfigure[Off-the-shelf IoT devices]{
\begin{minipage}{\linewidth}
\centering
\includegraphics[width=.9\textwidth]{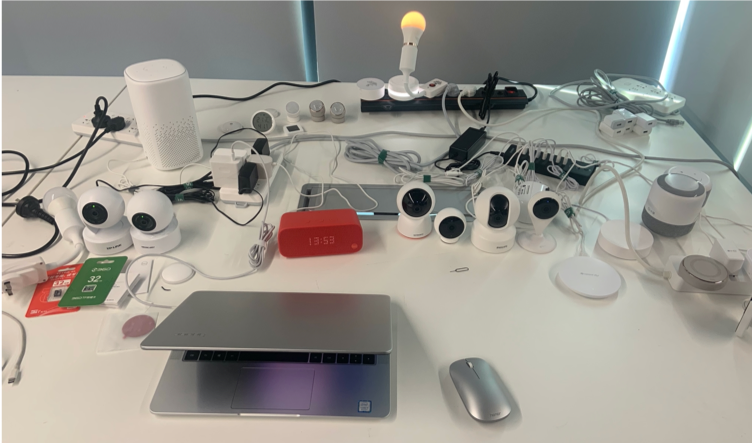}
\end{minipage}
}
\subfigure[Access point, collector and middlebox]{
\begin{minipage}{.9\linewidth}
\centering
\includegraphics[width=\textwidth]{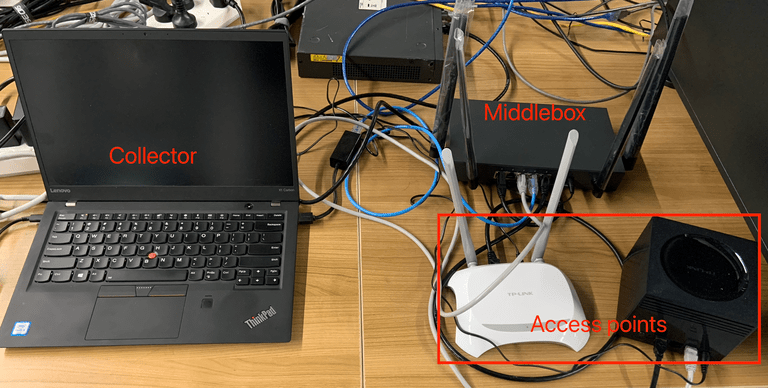}
\end{minipage}
}
\caption{IoT testbed photos.}
\label{fig:testbed_photo}
\end{figure}

For the idle/active experiment, to continuously activate the device functions, we employ an Android emulator installed with the IoT apps of the devices in the testbed to trigger their functions.
To automate this process, we use the Android Debug Bridge (ADB) that allows to monitor and control of the emulator. With this tool, we manually analyze the UI of the apps and record the operations to trigger a function. For example, tapping the app of a camera at certain pixel coordinates will start the live streaming. Then we can write a Bash script to for each app to automatically trigger its recorded functions. Listing~\ref{list:bash} gives an example of activating the live streaming of a Philips camera. We make the script run for 10 days for traffic collection.

The controlling scripts and part of our IoT traces in PCAP files have been released on \url{https://github.com/Ruoyu-Li/IoT_traffic_dataset}. 
For the complete dataset, please send an email to \url{huangyc20@mails.tsinghua.edu.cn}/\url{liry19@mails.tsinghua.edu.cn} for inquiry.

\begin{lstlisting}[caption=Sample Bash script to activate the live streaming of a Philips camera., label=list:bash, language=bash, mathescape]
# open app by tapping at pixel coordinates
((x=2d7));((y=554))
adb shell input tap x y
# wait for 20 seconds to guarantee app is open
adb shell sleep 20
# stream live for 5 minutes
((x=21c));((y=2bc))
adb shell input tap x y
adb shell sleep 300
# press home button and return to main page
adb shell input keyevent 3
adb shell sleep 1
# kill the app in the background
adb shell am force-stop com.philips.ipcamera
\end{lstlisting}

\newpage
\section{IoT Dataset Statistics}
\label{subsec:device_list}

\begin{table}[h]
    \caption{IoT dataset statistics (Self-built).}
    \footnotesize
    \centering
    \begin{tabular}{cccc}
    \toprule
    No. & Device & Num. of packets & Num. of flows \\
    \midrule
    0 & xiaomi-camera & 1362212 & 77408 \\
    1 & aqara-gateway & 148056 & 2 \\
    2 & skyworth-camera & 91284 & 422 \\
    3 & miai-soundbox & 1371918 & 136119 \\
    4 & honyar-outlet & 81450 & 1512 \\
    5 & hichip-camera & 33867 & 2025 \\
    6 & ihorn-gateway & 65436 & 154 \\
    7 & mercury-wirecamera & 363547 & 49435 \\
    8 & 360-doorbell & 466782 & 354 \\
    9 & ezviz-camera & 256319 & 16798 \\
    10 & philips-camera & 414182 & 1632\\
    11 & xiaomi-plug & 92436 & 157 \\
    12 & tmall-genie & 267424 & 2240 \\
    13 & gree-gateway & 22145 & 7 \\
    \bottomrule
    \end{tabular}
    \label{tab:full_device}
\end{table}

\begin{table}[h]
    \caption{IoT dataset statistics (UNSW).}
    \footnotesize
    \centering
    \begin{tabular}{cccc}
    \toprule
    No. & Device & Num. of packets & Num. of flows \\
    \midrule
    0 & Smart-Things & 391901 & 4904 \\
    1 & Amazon-Echo & 893171 & 58751 \\
    2 & Netatmo-Welcome & 402668 & 12537 \\
    3 & TP-Link-camera & 258557 & 2567 \\
    4 & Samsung-SmartCam & 258655 & 21555 \\
    5 & Withings-Baby-Monitor & 513013 & 9431 \\
    6 & TP-Link-plug & 22945 & 1071 \\
    7 & NEST-smoke-alarm & 3221 & 149 \\
    8 & Netatmo-weather-station & 38537 & 4024 \\
    9 & Withings-scale & 4238 & 3 \\
    10 & Blipcare-Blood-Pressure-meter & 124 & 6 \\
    11 & Withings-sleep-sensor & 273948 & 5738 \\
    12 & Triby-Speaker & 102369 & 330 \\
    13 & PIX-STAR-Photo-frame & 35409 & 2231 \\
    14 & HP-Printer & 41785 & 90 \\
    \bottomrule
    \end{tabular}
    \label{tab:full_device_UNSW}
\end{table}

\begin{table}[h]
    \caption{IoT dataset statistics (ICL).}
    \footnotesize
    \centering
    \begin{tabular}{cccc}
    \toprule
    No. & Device & Num. of packets & Num. of flows \\
    \midrule
    0 & samsungtv-wired & 91199 & 10220 \\
    1 & smartthings-hub & 42244 & 7476 \\
    2 & google-home & 55211 & 12237 \\
    3 & wansview-cam-wired & 230827 & 98452 \\
    4 & blink-security-hub & 20122 & 226 \\
    5 & netatmo-weather-station & 3864 & 330 \\
    6 & wemo-plug & 1392 & 8 \\
    7 & bosiwo-camera-wifi & 2175 & 34 \\
    8 & philips-hub & 9209 & 572 \\
    9 & lightify-hub & 16612 & 10 \\
    10 & tplink-bulb & 5887 & 378 \\
    11 & sengled-hub & 11568 & 336 \\
    12 & sousvide & 494631 & 60 \\
    13 & tplink-plug & 1345 & 114 \\
    14 & charger-camera & 2854 & 559 \\
    15 & xiaomi-cam & 146335 & 30 \\
    16 & ring-doorbell & 3994 & 42 \\
    17 & magichome-strip & 33725 & 9 \\
    18 & smarter-coffee-mach & 3992 & 6 \\
    19 & honeywell-thermostat & 3026 & 2 \\
    20 & nest-tstat & 29777 & 1784 \\
    \bottomrule
    \end{tabular}
    \label{tab:full_device_ICL}
\end{table}

\newpage

\begin{table}[h]
    \caption{IoT dataset statistics (NEU).}
    \footnotesize
    \centering
    \begin{tabular}{cccc}
    \toprule
    No. & Device & Num. of packets & Num. of flows \\
    \midrule
    0 & tplink-bulb & 6155 & 67 \\
    1 & lightify-hub & 3514 & 8 \\
    2 & google-home-mini & 57577 & 12764 \\
    3 & yi-camera & 23522 & 492 \\
    4 & appletv & 8418 & 530 \\
    5 & smartthings-hub & 57135 & 7772 \\
    6 & wansview-cam-wired & 240919 & 98943 \\
    7 & magichome-strip & 4505 & 18 \\
    8 & amazon-echodot & 38702 & 6117 \\
    9 & blink-security-hub & 18536 & 405 \\
    10 & nest-tstat & 11289 & 586 \\
    11 & wemo-plug & 10002 & 162 \\
    12 & amcrest-cam-wired & 22358 & 4663 \\
    13 & behmor-brewer & 31091 & 10 \\
    14 & xiaomi-bulb & 10229 & 14 \\
    15 & amazon-cloudcam & 611172 & 2173 \\
    16 & samsung-fridge & 49254 & 10407 \\
    17 & smarter-ikettle & 5266 & 4 \\
    18 & insteon-hub & 47539 & 1020 \\
    19 & cortana-invoke & 99935 & 2057 \\
    20 & lefun-cam-wired & 6976 & 75 \\
    21 & microseven-camera & 10392 & 114 \\
    22 & ge-microwave & 37997 & 156 \\
    23 & sengled-hub & 11076 & 11 \\
    24 & wink-hub & 17022 & 230 \\
    25 & zmodo-doorbell & 10379 & 52 \\
    \bottomrule
    \end{tabular}
    \label{tab:full_device_NEU}
\end{table}